\documentclass[11pt]{article}
\usepackage{fancyheadings, bbm, bm}
\usepackage{natbib,comment,tabularx}
\usepackage{color}
\makeatletter
\def\section{\@startsection{section}{1}
{\z@}{-2.5ex plus -0.5ex minus -0.1ex}{0.5ex plus 0.1ex}{\centering\large\bfseries}}
\def\subsection{\@startsection{subsection}{2}
{\z@}{-2.25ex plus -0.3ex minus -0.1ex}{0.05ex plus 0.05ex}{\normalsize\bfseries}}
\def\subsubsection{\@startsection{subsubsection}{3}
{\z@}{-2.25ex plus -0.3ex minus -0.2ex}{0.05ex plus 0.05ex}{\normalsize\bfseries\rm}}
\makeatother
\usepackage{amsmath, amsfonts,amstext,rotating,amssymb,graphics,xspace, setspace}
\usepackage{amsthm}
\usepackage{epsfig}
\usepackage[retainorgcmds]{IEEEtrantools}
\usepackage{hyperref}
\hypersetup{
  colorlinks   = true, 
  urlcolor     = blue, 
  linkcolor    = blue, 
  citecolor   = blue    
}
\usepackage [left=1in,right=1in, top=1in, bottom=1in] {geometry}
\newcommand{\ba}{\begin{array}}
\newcommand{\ea}{\end{array}}
\newcommand{\bs}{\begin{align}\begin{split}\nonumber}
\newcommand{\bsnumber}{\begin{align}\begin{split}}
\newcommand{\es}{\end{split}\end{align}}

\newcommand{\bi}{\begin{itemize}}
\newcommand{\ei}{\end{itemize}}
\newcommand{\be}{\begin{enumerate}}
\newcommand{\ee}{\end{enumerate}}

\def\PP{\mathbb{P}}    
\def\EE{\mathbb{E}}    
\def\G{{\mathcal G}}   

\theoremstyle{definition}
\newtheorem{example}{Example}

\theoremstyle{plain}
\newtheorem{thm}{Proposition}

\newtheorem{assumption}{Assumption}

\begin{document}
\lhead{}
\rhead{\thepage}
\cfoot{}
\cfoot{\fancyplain{}}

\author{Emerson Melo\thanks{Department of Economics, Indiana University Bloomington. Email: {\tt  emelo@iu.edu}} \and Kirill Pogorelskiy \thanks{Department of Economics, University of Warwick. Email: {\tt k.pogorelskiy@warwick.ac.uk}} \and Matthew Shum\thanks{Division of the Humanities and Social Sciences, California Institute of Technology. Email:  {\tt mshum@caltech.edu}}}
\title{Testing the Quantal Response Hypothesis\thanks{Acknowledgments:
    We thank Marina Agranov, Larry Blume, Federico Echenique, Ben Gillen, Russell Golman,  Phil Haile, J\"{o}rg Stoye, Leeat Yariv and especially Tom Palfrey for support and insightful comments. We also thank the participants of the Yale conference on Heterogenous Agents and Microeconometrics. We are grateful to Michael McBride and Alyssa Acre at UC Irvine ESSL laboratory for help with running experiments.}}
\date{\noindent {\small \today}
\\}
\maketitle

\begin{abstract}
This paper develops a non-parametric test for consistency of players' behavior in a series of games with the Quantal Response Equilibrium (QRE). 
The test exploits a characterization of the equilibrium choice probabilities in \emph{any} structural QRE as the gradient of a
convex function, which thus satisfies the {\em cyclic monotonicity} inequalities.  Our testing procedure
utilizes recent econometric results for moment inequality models.
We assess our test using lab experimental data from a series of generalized matching pennies games.
We reject the QRE hypothesis in the pooled data, but it cannot be rejected in the individual data for over half of the subjects.


\end{abstract}

\section{Introduction}
A vast literature in experimental economics has demonstrated that, across a wide variety of games, behavior deviates systematically from Nash equilibrium-predicted behavior.
%
In order to relax Nash equilibrium in a natural fashion, while preserving the idea of equilibrium, \citet{MP95} introduced the notion of {\em Quantal Response Equilibrium (QRE)}.   QRE has becine a popular tool in experimental economics because typically it provides an improved fit to the experimental data; moreover, it is also a key model in behavioral game theory and serves as a benchmark for tractable alternative theories of bounded rationality.\footnote{See \citet{Camerer03} and \citet{CrawfordEtAl13}.
The latest book-length treatment of the theory behind QRE and its numerous applications in economics and political science is forthcoming in \citet{GoereeHoltPalfrey16}.}
While most existing work estimating structural QRE (e.g., \citet{MerloPalfrey13}, \citet{CamererEtAl15} to name a few) utilize the \emph{parametric}, logit version of QRE and its variants, the \emph{nonparametric} tests of QRE have not been available.

This paper is the first to develop and implement a formal nonparametric procedure to test, using experimental data, whether subjects are indeed behaving according to QRE.  Our approach is related to the econometrics literature on semiparametric discrete choice models and moment inequalities.
Our test is based on the notion of {\em cyclic monotonicity}, a concept from convex analysis which is useful as a characterizing feature of convex potentials.   Cyclic monotonicity imposes joint inequality restrictions between the underlying choice frequencies and payoffs from the underlying games; hence, we are able to apply tools and methodologies from the recent econometric literature on moment inequality models to derive the formal statistical properties of our test.  Importantly, our test for QRE is  {\em nonparametric} in that one need not specify a particular probability distribution for the random shocks; thus, the results are robust to a wide variety of distributions.

Subsequently, we apply our test to data from a lab experiment on generalized matching pennies games.
We find that QRE is rejected soundly when data is pooled across all subjects and all plays of each game.   But when we consider subjects individually, we find that the QRE hypothesis cannot be rejected for upwards of half the subjects.   This suggests that there is substantial heterogeneity in behavior across subjects.   Moreover, the congruence of subjects' play with QRE varies substantially depending on whether subjects are playing in the role of the Row vs. Column player.

Our work here builds upon and extends \citet{HaileHortacsuKosenok08} (hereafter HHK) who showed that, that without imposing strong assumptions on the shock distributions, QRE can rationalize any outcome in a given game.
HHK describe several approaches for testing for QRE, but did not provide guidance for formal econometric implementation of such tests. We build upon one of these approaches, based on the relation between changes in QRE probabilities across the series of games that only differ in the payoffs and the changes in the respective expected payoffs, and develop an econometric test for consistency of the data with a QRE in this more general case.

Our use of the notion of cyclic monotonicity to test the QRE hypothesis
appears new to the 
literature. Elsewhere, cyclic
monotonicity has been studied in the context of multidimensional mechanism
design. In particular, the papers  by \citet{Rochet87}, \citet{SaksYu05},
\citet{LaviSwamy09}, \citet{Ashlagietal10}, and \citet{ArcherKleinberg14} (summarized in \citet[Chapter 4]{Vohrabook}), relate the incentive compatibility (truthful implementation) of a
mechanism to its cyclic monotonicity properties. Similarly,  the papers by \cite{FosgeraudePalma15} and \citet{McFaddenFosgerau12} introduce cyclic monotonicity
to study revealed preference in discrete choice models.


Finally, in this paper we apply the cyclic monotonicity to test for QRE using experimental data in which players' payoffs are known (because they are set by the experimenter).  The cyclic monotonicity property may also be useful in settings where the researcher does not completely observe agents' utilities.   In other work involving one of the authors \citep{ShiShumSong15} we have also used cyclic monotonicity for identification and estimation of semiparametric multinomial choice models, in which the payoff shocks are left unspecified (as here), but the utility functions are parametric and assumed to be known up to a finite-dimensional parameter.


The rest of the paper is organized as follows. Section \ref{background} presents the QRE approach. Section \ref{test} introduces the test for the QRE hypothesis, and Section \ref{sec:moment_inequalities} discusses the moment inequalities for testing. Section \ref{stat_properties} discusses the statistical properties of the test.  Section \ref{exp_design} describes our experiment, with subsections \ref{subsec:exp_design}
and \ref{exp_results} presenting the experimental design and results respectively. Section \ref{concl_exte} concludes.
Appendix \ref{CM_UM} provides additional details about the interpretation of the cyclic monotonicity inequalities.
Appendices \ref{proof:prop1} and \ref{proof:prop:qreUniq} contain omitted proofs and additional theoretic results. Appendix \ref{appxC} contains omitted computational details of the test.
Appendix \ref{appx} contains experimental instructions.

\section{QRE background}\label{background}
In this section we briefly review the main ideas behind the QRE approach.
We use the notation from \citet{MP95}.

Consider a finite $n$-person game $G(N, \{S_i\}_{i\in N}, \{u_i\}_{i\in N})$.
The set of pure strategies (actions) available to player $i$ is indexed by $j=1,\ldots, J_i$, so that $S_i=\{s_1,\ldots, s_{J_i}\}$, with a generic element denoted $s_{ij}$.
Let $\bm{s}$ denote an $n$-vector strategy profile;  let $s_i$ and $\bm{s}_{-i}$ denote player $i$'s (scalar) action and the vector of actions for all players other than $i$. In terms of notation, all vectors are denoted by bold letters.
Let $p_{ij}$ be the probability that player $i$ chooses action $j$, and $\bm{p}_i$ denote the vector of player $i$'s choice probabilities. Let $\bm{p}=(\bm{p}_1,\ldots, \bm{p}_n)$ denote the vector of probabilities across all the players. Player $i$'s utility function is given by $u_i(s_i, \bm{s}_{-i})$. At the time she chooses her action, she does not know what actions the other players will play.
Define the expected utility that player $i$ gets from playing a pure strategy $s_{ij}$ when everyone else's joint strategy is $\bm{p}_{-i}$ as
\begin{equation*}
u_{ij}(\bm{p}) \equiv u_{ij}(\bm{p}_{-i})=\sum_{\bm{s}_{-i}} p(\bm{s}_{-i}) u_i(s_{ij}, \bm{s}_{-i}),
\end{equation*}
where $\bm{s}_{-i}=(s_{k{j_k}})_{k\in N_{-i}}$, and $p(\bm{s}_{-i})=\prod_{k\in N_{-i}} p_{k{j_k}}$.

In the QRE framework uncertainty is generated by players' making ``\emph{mistakes}''.
This is modelled by assuming that, given her beliefs about the opponents' actions $\bm{p}_{-i}$, when choosing her
action, player $i$ does not choose the action $j$ that maximizes her expected utility $u_{ij}(\bm{p})$, but rather chooses the action that maximizes $u_{ij}(\bm{p})+\varepsilon_{ij}$, where $\varepsilon_{ij}$ represents a preference shock at action $j$.
For each player $i\in N$ let $\bm{\varepsilon}_i=(\varepsilon_{i1},\ldots,\varepsilon_{iJ_i})$ be drawn according to an absolutely continuous distribution $F_i$ with mean zero.\footnote{Notice that as a result each action can be chosen with a positive probability, ruling out consistency with a pure-strategy Nash equilibrium (which can be restored in the limit as the shocks go to zero.) This is never a concern for our test since the choice probabilities are estimated from the data and 
necessarily contain some noise ruling out pure strategies. Furthermore all games in our application have a unique totally-mixed Nash equilibrium. }  Then an expected utility maximizer, player $i$, given beliefs $\bm{p}$, chooses action $j$ iff
$$
u_{ij}(\bm{p})+\varepsilon_{ij} \geq u_{ij'}(\bm{p})+\varepsilon_{ij'}, \quad \forall j'\neq j.
$$
Since preference shocks are random, the probability of choosing action $j$ given beliefs $\bm{p}$, denoted $\pi_{ij}(\bm{p})$,  can be formally expressed as
\begin{IEEEeqnarray}{rCl}
\pi_{ij}(\bm{p})&\equiv& \PP\left(j =\arg\max_{j'\in\{1,\ldots,J_i\}} \left\{u_{ij'}(\bm{p})+\varepsilon_{ij'}\right\}\right) \nonumber \\
&=&\int_{\left\{\bm{\varepsilon}_i \in \mathbb{R}^{J_i}|\ u_{ij}(\bm{p})+\varepsilon_{ij} \geq u_{ij'}(\bm{p})+\varepsilon_{ij'}\ \forall j'\in\{1,\ldots, J_i\}\right\}} dF_i(\bm{\varepsilon}_i) \label{qre_p}
\end{IEEEeqnarray}
Then a \emph{Quantal Response Equilibrium} is defined as a set of choice probabilities $\left\{\pi^*_{ij}\right\}$ such that for all $(i,j)\in N\times \{1,\ldots,J_i\}$,
$$
\pi^*_{ij}=\pi_{ij}(\bm{\pi^*})
$$
Throughout, we assume that all players' preference shock distributions
are \emph{invariant}; that is, the distribution does not depend on the
payoffs:
\begin{assumption}\label{assumption:invariance}
(Invariant shock distribution) For all realizations $\bm{\varepsilon}_i:=(\varepsilon_{i1},\ldots,\varepsilon_{i{J_i}})$ and all payoff functions $u_i(\cdot)$, we have $F_i(\bm{\varepsilon}_i|u_i(\cdot))=F_i(\bm{\varepsilon}_i)$.
\end{assumption}
Such an invariance assumption was also considered in
HHK's study of the quantal response model, and also assumed in most empirical implementations of QRE.\footnote{
Most applications of QRE assume that the utility shocks follow a logistic distribution, regardless of the magnitude of payoffs.  One exception is \citet{McKelveyPalfreyWeber00}, who allow the logit-QRE parameter to vary across different games. This direction is further developed in \citet{RogersPalfreyCamerer09}.}
In the specific environment of our test here, the invariance assumption allows us to compare choice probabilities across different tests using the property of {\em cyclic monotonicity}, which we explain in the next section.

\section{A test based on convex analysis}\label{test}

In this section we propose a test for the QRE hypothesis. We start by defining the following function:
\begin{equation}
\varphi^i(\bm {u}_{i}(\bm{\pi}))\!\equiv \!\EE\left[\max_{j\in S_i}\{u_{ij}(\bm{\pi})\!+\!\varepsilon_{ij}\}\right] \label{eq:phi}
\end{equation}

In the discrete choice model literature, the expression $\varphi(\bm{u})$ is known as the social surplus function.\footnote{For details see \citet{McFadden81}. } Importantly, this function is smooth and convex.
Now the QRE probabilities $\pi_{ij}(\bm{\pi}^{*})$ can be expressed as
\begin{equation}
  \bm{\pi}_{i}^*=\!\bm{\nabla}\varphi^i(\bm{u}_{i}(\bm{\pi^*})) \label{cvx_rep}
\end{equation}
This follows from the well-known Williams-Daly-Zachary theorem from discrete-choice theory (which can be considered a version of Roy's Identity for discrete choice models; see \citet[p.3104]{Rust94}).
Thus if Eq. \eqref{cvx_rep} holds for all players $i$, then $\left\{\pi^*_{ij}\right\}$  is a quantal response equilibrium. 
Eq. (\ref{cvx_rep}) characterizes the QRE choice probabilities as the gradient of the convex function $\varphi$.  It is well-known \citep[Theorem 24.8]{Rockafellar70} that the gradient of a convex function satisfies a {\em cyclic   monotonicity} property. This property is the generalization, for functions of several variables, of the fact that the derivative of a univariate convex function is monotone nondecreasing.

To define cyclic monotonicity in our setting, consider a {\em cycle}\footnote{A \emph{cycle} of length $\mathcal{L}$ is just a sequence of $\mathcal{L}$ games $G_0,\ldots,G_{\mathcal{L}-2}, G_{\mathcal{L}-1}$ with $G_{\mathcal{L}-1}=G_0$.} of games $C\equiv \left\{ G_0,G_1,G_2,\ldots,G_0\right\}$ where $G_m$ denotes the game at index $m$ in a cycle.   These games are characterized by the same set of choices for each player, and the same distribution of payoff shocks (i.e., satisfying Assumption \ref{assumption:invariance}), but distinguished by payoff differences. Let $[\bm{\pi}^*_i]^m$ denote the QRE choice probabilities for player $i$ in game $G_m$, and $\bm{u}_i^m\equiv \bm{u}_i^m([\bm{\pi}^*]^m)$ the corresponding equilibrium expected payoffs.
Then the cyclic monotonicity property says that
\begin{equation}\label{basicCM}
\sum_{m=G_0}^{G_{\mathcal{L}-1}} \bigg< [\bm{u}_i]^{m+1}-[\bm{u}_i]^{m}, [\bm{\pi}^*_i]^m \bigg>\leq 0
\end{equation}
for all finite cycles of games of length $\mathcal{L}\geq 2$, and all players $i$.\footnote{
Under convexity of $\varphi^i(\cdot)$, we have $\varphi^i(\bm{u}_i^{m+1})\geq \varphi^i(\bm{u}_i^{m}) + \big<\nabla\varphi^i(\bm{u}_i^m), (\bm{u}_i^{m+1}-\bm{u}_i^m)\big>$.  Substituting in $\nabla\varphi^i(\bm{u}_i^m) = \bm{\pi}_i^m$ and summing across a cycle, we obtain the CM inequality in Eq. (\ref{basicCM}).
} Expanding the inner product notation, the Cyclic Monotonicity (CM) conditions may be written as follows:
\begin{equation}
\sum_{m=G_0}^{G_{\mathcal{L}-1}} \sum_{j=1}^{J_i} \left(u_{ij}^{m+1}-u_{ij}^{m}\right) [\pi^*_{ij}]^m \leq 0 \label{cm_test}
\end{equation}
This property only holds under the invariance assumption.  Without it, the payoff shock distributions, and hence the social surplus functions, will be different across each game,  implying that their corresponding choice probabilities are gradients of different social surplus functions.  Because of this, the cyclic monotonicity property need not hold.

 The number of all finite game cycles times the number of players can be, admittedly, very large. To reduce it, we note that
the cyclic monotonicity conditions \eqref{cm_test} are invariant under the change of the starting game index in the cycles; for instance, the inequalities emerging from the cycles $\left\{G_i,\ldots,G_j, G_k,G_i\right\}$ and $\left\{G_j,\ldots,G_k, G_i,G_j\right\}$ are the same.

Intuitively  CM conditions can be derived from, and also imply, players' utility maximization in appropriately perturbed games. In particular, in Appendix \ref{CM_UM} we  prove the following result
\begin{thm}\label{prop_CM_UM}
 Consider a cycle of length $\mathcal{L}\geq2$. The cyclic monotonicity conditions hold if and only if each player's choice probabilities in each game along the cycle maximize the difference in her expected utility between every two adjacent games in the cycle.
\end{thm}
Proposition \ref{prop_CM_UM} establishes the equivalence between the cyclic monotonicity condition and utility maximization among a family of perturbed games that only vary in the payoffs.   
The equivalence in proposition critically depends on the fact that equilibrium probabilities are given by expression \eqref{cvx_rep}.  Formally, expression \eqref{cvx_rep} allows us to use  \emph{conjugate duality} arguments to show that  CM is equivalent to players' utility maximization. Thus Proposition \ref{prop_CM_UM}  helps us interpret our test directly in terms of the main QRE property of positive responsiveness, where each action's probability is increasing in the action's expected payoff.


\textbf{ Remark: Cyclic monotonicity and incentive compatibility.}
Proposition  \ref{prop_CM_UM}  also implies that the CM inequalities can be interpreted as ``incentive compatibility'' conditions on players' choices across games. Namely, if the CM inequalities are violated for player $i$ and some cycle of games $C$, then $i$ is not optimally adjusting her choice probabilities in response to changes in the expected payoffs across the games in the cycle $C$. To see this, consider firstly a family of games with only two actions, which vary in the expected payoff difference $\mathbb{E}(u_2-u_1)$ between actions 2 and 1.   Obviously, utility maximization should imply that the choice probabilities of playing action 2 across these games should be nondecreasing in the expected payoff differences; alternatively, monotonicity in expected payoff differences is an ``incentive compatibility'' condition on choice probabilities across these games.   

In games with more than two actions, the expected payoff differences (relative to a benchmark action) form a vector, as do the choice probabilities for each action.   Proposition  \ref{prop_CM_UM}  and the discussion above show that utility maximization across a family of games varying in expected payoff differences imply that the vector of corresponding choice probabilities in each game can be represented as a gradient of the social surplus function; that is, cyclic monotonicity is an incentive compatibility condition on choice probabilities in a family of games with more than two actions.\hfill$\blacksquare$

\textbf{ Special case: Two actions.}
As the previous remark pointed out, in a family of games with only two actions, cyclic monotonicity reduces to the usual monotonicity.   That is, in these games, the cyclic monotonicity conditions \eqref{cm_test} only need to be checked for cycles of length $2$.\footnote{Formally, this fact follows from the observation that for games with two actions, we can rewrite the functions $\varphi^i(u_i(\bm{\pi}))$ as $\varphi^i(u_{i1}(\bm{\pi}))=\varphi^i(u_{i1}(\bm{\pi})-u_{i2}(\bm{\pi}),0)+u_{i2}(\bm{\pi})$. Since without loss of generality we can normalize $u_{i2}(\bm{\pi})$ to be constant, we obtain that $\varphi^i(u_i(\bm{\pi}))$ is a univariate function. Using \citet[Proposition 2]{Rochet87} we conclude that if $\varphi^i$  satisfies  \eqref{cm_test} for all cycles of length $2$, then \eqref{cm_test} is also satisfied for cycles of arbitrary length $\mathcal{L}>2$. }
Because many experiments study games where players' strategy sets consist of two elements, this observation turns out to be  useful from an applied perspective.\hfill$\blacksquare$


\section{Moment inequalities for testing cyclic monotonicity}\label{sec:moment_inequalities}
Consistency with QRE can be tested nonparametrically from experimental data in which the same subject $i$ is playing a series of one-shot games with the same strategy spaces such that each game is played multiple times.
In this case, the experimental data allows to estimate a vector of probabilities $[\bm {\pi}^*_i]^m \in \Delta(S_i)$ for each game $m$ in the sample, and we can compute the corresponding equilibrium expected utilities $[\bm {u}_i]^m$ (assuming risk-neutrality).

Suppose there are $M\geq 2$ different games in the sample. We assume that we are able to obtain estimates of $\hat{\bm{\pi}}_i^m$, the empirical choice frequencies, from the experimental data, for each subject $i$ and for each game $m$. Thus we compute
$\hat{\pi}_{ij}^m$ from $K$ trials for subject $i$ in game $m$:
$$\hat{\pi}_{ij}^m=\frac{1}{K} \sum_{k=1}^K \mathbbm{1}_{\{\text{$i$ chooses $j$
    in trial $k$ of game $m$}\}}$$
This will be the source of the sampling error in our econometric setup.   Also, let $\hat{\bm{u}}_i^m
\equiv\bm{u}_i^m(\hat{\bm{\pi}}^m)$ be the estimated equilibrium expected utilities
obtained by plugging in the observed choice probabilities
$\hat{\bm{\pi}}^m$ into the payoffs in game $m$. Then the sample moment inequalities take the following form: for all cycles of length $\mathcal{L}\in\{2,\ldots, M\}$
\begin{equation}
\sum_{m=G_0}^{G_{\mathcal{L}-1}} \sum_{j=1}^{J_i}
\left(\hat{u}_{ij}^{m+1}-\hat{u}_{ij}^{m}\right) \hat {\pi}_{ij}^m \label{cm_sample_test}
\leq 0
\end{equation}
Altogether, in an $n$-person game we have $n \sum_{\mathcal{L}=2}^{M}\#C(\mathcal{L})$ moment inequalities, where $\#C(\mathcal{L})$ is the number of different (up to a change in the starting game index) cycles of length $\mathcal{L}$. These inequalities make up a necessary condition for a finite sample of games to be QRE-consistent.

\subsection{``Cumulative rank'' test as a special case of Cyclic Monotonicity}

HHK propose alternative methods of testing
the QRE model based on cumulative rankings of choice probabilities across perturbed games,\footnote{HHK also consider testing the QRE hypothesis using ``Block-Marschak'' polynomials.  QRE play implies linear inequalities involving the Block-Marschak polynomials which may be tested formally using similar methods as we describe in this paper.  However, the two approaches are qualitatively quite different as the Block-Marschak approach involves comparing games which vary in the actions available to players and, as HHK point out, can be feasibly tested in the lab only for special families of games (such as Stackelberg games or games with some nonstrategic players).   In contrast,
our approach, based on cyclic monotonicity, compares games which vary in players' payoffs.   Hence, for these reasons, a full exploration of the Block-Marschak approach seems beyond the scope of this paper.} which also imply stochastic equalities or inequalities involving estimated choice probabilities from different games.
We will show here that, in fact, our CM conditions are directly
related to HHK's rank-cumulative probability conditions in the special
case when there are only two games (i.e. all cycles are of length 2), and under a
certain non-negativity condition on utility differences between the games.

Formally, HHK consider two perturbed games with the same strategy spaces and re-order strategy indices for each player $i$ such that
\begin{IEEEeqnarray*}{rCl}
\tilde{u}_{i1}^1-\tilde{u}_{i1}^0 \geq \tilde{u}_{i2}^1-\tilde{u}_{i2}^0 \geq \ldots \geq \tilde{u}_{iJ_i}^1-\tilde{u}_{iJ_i}^0
\end{IEEEeqnarray*}
where $ \tilde{u}_{ij}^m\equiv u_{i}(s_{ij},\bm{\pi}^m_{-i})-\frac{1}{J_i}\sum_{j=1}^{J_i}u_{i}(s_{ij},\bm{\pi}^m_{-i})$ for $m=0,1$. These inequalities can be equivalently rewritten as
\begin{IEEEeqnarray}{rCl}
u_{i1}^1- u_{i1}^0 \geq u_{i2}^1- u_{i2}^0 \geq \ldots \geq u_{iJ_i}^1-u_{iJ_i}^0 \label{cond1}
\end{IEEEeqnarray}
HHK's Theorem $2$ states that given the indexing in \eqref{cond1} and assuming Invariance (see Assumption \ref{assumption:invariance}), QRE consistency implies the following {\em cumulative rank property:}
\begin{IEEEeqnarray}{rCl}
\sum_{j=1}^k( \pi_{ij}^1-\pi_{ij}^0)\geq 0 \quad
\text{for all $k=1,\ldots, J_i$.}\label{rcTestHHK}
\end{IEEEeqnarray}
 This property is related to our test as the following proposition demonstrates (the proof is in Appendix \ref{proof:prop1}):
\begin{thm}\label{prop1}
 Let $M=2$. If all expected utility differences in \eqref{cond1} are non-negative, then HHK's cumulative rank condition \eqref{rcTestHHK} implies the CM inequalities \eqref{cm_test}. Conversely, the CM inequalities \eqref{cm_test} imply the cumulative rank condition \eqref{rcTestHHK} (without additional assumptions on expected utility differences).
\end{thm}
Hence for the special case of just two games HHK's cumulative rank property can be directly related to the cyclic monotonicity inequalities.
More broadly, Theorem \ref{prop1} implies that the cumulative rank testing approach is similar to testing cyclic monotonicity using only length-2 cycles (ie. using only pairwise comparisons among games).  It is known that, when there are more than 2 games (with more than 2 strategies in each game), the pairwise comparisons do not exhaust the restrictions in the cyclic monotonicity inequalities.  (See \citet{SaksYu05}, \citet{Ashlagietal10}, and \citet[Chapter 4]{Vohrabook}.)

\subsection{Limitations and extensions of the test} \label{sec:test_limitations}

Our test makes use of expected payoffs and choice probabilities in a fixed set of $M\geq 2$ games.

To estimate expected payoffs we had to assume that players are risk-neutral. This assumption might be too strong a priori (e.g., \citet{GoereeHoltPalfrey00} argue that risk aversion can help explain QRE inconsistencies). Notice, however, that the test itself does not depend on risk-neutrality: it only requires that we know the form of the utility function. Thus under additional assumptions about the utility, we can also investigate how risk aversion affects the test results. See Section \ref{exp_results} for details.


Our test also assumes that for each of the games considered,
there is only one unique QRE. Note that since we do not specify the
distribution of the random utility shocks, this uniqueness assumption
is not verifiable.
However, as in much of the recent empirical games literature in industrial organization\footnote{
See, e.g., \citet{AguirregabiriaMira07}, \citet{Bajarietal2007}.}, our testing procedure strictly speaking only assumes that {\em there is a unique equilibrium played in the data}.\footnote{Indeed, practically all of the empirical studies of experimental data utilizing the quantal response framework assume that a unique equilibrium is played in the data, so that the observed choices are drawn from a homogeneous sampling environment.  For this reason, our test may not be appropriate for testing for QRE using field data, which were not generated under these controlled laboratory experimental conditions. See \citet{DePaulaTang2012} for a test of multiple equilibria presence in the data.}
Several considerations make us feel that this is a reasonable assumption in our application.
First, in our experiments, the subjects are randomly matched across different rounds of each game, so that playing multiple equilibria in the course of an experiment would require a great deal or coordination.
Second, as we discuss below, and in Appendix \ref{proof:prop:qreUniq}, all the experimental games that we apply our test to in this paper have a unique QRE under an additional regularity assumption.

Notwithstanding the above discussion, in some potential applications our test may wrongly reject (i.e., it is biased) the QRE null hypothesis when there are multiple quantal response equilibria played in the data. Given the remarks here, our test of QRE should be generally considered a joint test of the QRE hypothesis along with those of risk neutrality of the subjects, invariant shock distribution (Assumption \ref{assumption:invariance}), and unique equilibrium in the data.

\section{Econometric implementation: Generalized moment selection procedure}\label{stat_properties}
In this section we consider the formal econometric properties of our test, and the application of the generalized model selection procedure of \citet{AndrewsSoares10}.
Let $\bm{\nu}\in\mathbb{R}^{P}$ denote the vector of the left hand sides of the cyclic monotonicity inequalities \eqref{cm_test}, written out for all cycle lengths and all players. Here $P\equiv n \sum_{\mathcal{L}=2}^{M}\#C(\mathcal{L})$ and $\#C(\mathcal{L})$ is the number of different (up to the starting game index) cycles of length $\mathcal{L}$. Let us order all players and all different cycles of length $\mathcal{L}$ from $2$ to $M$ in a single ordering, and for $\ell\in\{1,\ldots,P\}$, let $\mathcal{L}(\ell)$ refer to the cycle length at coordinate number $\ell$ in this ordering, $m_0(\ell)$ refer to the first game in the respective cycle, and $\iota(\ell)$ refer to the corresponding player at coordinate number $\ell$.
Then we can write $\bm{\nu}\equiv(\nu_{1},\ldots,\nu_\ell, \ldots, \nu_{P})$ where each generic component $\nu_\ell$ is given by \eqref{cm_sample_test}, i.e.
\begin{IEEEeqnarray}{rCl}
\nu_\ell&=&\sum_{m=m_0(\ell)}^{\mathcal{L}(\ell)-1}\sum_{j=1}^{J_{\iota(\ell)}} \sum_{\bm{s_{-\iota(\ell)}}} \pi_{\iota(\ell)j}^m \left(\left(\prod_{k\in N_{-\iota(\ell)}} \pi_{k_{j_k}}^{m+1}\right)u_{\iota(\ell)}^{m+1}(s_{\iota(\ell)j},\bm{s_{-\iota(\ell)}})\right.\nonumber \\
&-&\left.\left(\prod_{k\in N_{-\iota(\ell)}} \pi_{k_{j_k}}^{m}\right)u_{\iota(\ell)}^{m}(s_{\iota(\ell)j},\bm{s_{-\iota(\ell)}})\right)
  \label{cm_test_explicit}
\end{IEEEeqnarray}

Define $\bm{\mu}\equiv -\bm{\nu}$, then cyclic monotonicity is equivalent to $\bm{\mu} \geq \bm{0}$.
Let $\bm{\hat\mu}$ denote the estimate of $\bm{\mu}$ from our experimental data.
In our setting, the sampling error is in the choice probabilities
$\pi$'s. Using the Delta method, we can derive that, asymptotically (when the number of trials of each game out of a fixed set of $M$ games goes to infinity),
$$
\bm{\hat\mu} \overset{a}{\sim} N(\bm{\mu_0}, \Sigma)\quad\text{and }\ \Sigma=JV J'
$$
where $V$ denotes the variance-covariance matrix for the $Mn\times
1$-vector $\bm{\pi}$ and $J$ denotes the $P\times Mn$ Jacobian matrix
of the transformation from $\bm{\pi}$ to $\bm{\mu}$.   Since $P>>Mn$,
the resulting matrix $\Sigma$ is singular.\footnote{Note also that $\Sigma$ is
the approximation of the {\em finite-sample} covariance matrix, so
that the square-roots of its diagonal elements correspond to the
standard errors;
i.e. the elements are already ``divided through'' by the sample size, which accounts for the differences between the equations below and the corresponding ones in \citet{AndrewsSoares10}.}


We perform the following hypothesis test:
\begin{equation}\label{hypotheses}
H_0:\ \bm{\mu_0}\geq \bm{0}\quad \textit{vs.} \quad H_1:\ \bm{\mu_0}\not\geq \bm{0},
\end{equation}
where $\bm{0}\in\mathbb{R}^P$.
Letting $\hat\Sigma$ denote an estimate of $\Sigma$, we utilize the following test statistic
\begin{equation}
S(\mathbf{\hat\mu},\hat\Sigma):=\sum_{\ell=1}^P \left[ \hat\mu^\ell/\hat\sigma_\ell\right]_{-}^2
\label{iu_test_stat}\end{equation}
where $[x]_{-}$ denotes $x\cdot \mathbbm{1}(x<0)$, and
$\hat\sigma_1^2,\ldots, \hat\sigma_P^2$ denote the diagonal elements
of $\hat\Sigma$.   The test statistic is sum of squared violations
across the moment inequalities, so that larger values of the statistic indicate evidence against the null hypotheses (\ref{hypotheses}).   Since there are a large number of moment inequalities in Eq. (\ref{hypotheses}) (in the
application, $P=40$), we
utilize the Generalized Moment Selection (GMS) procedure of
\citet{AndrewsSoares10} (hereafter ``AS'').
This procedure combines moment selection along with hypothesis testing, and is especially useful when there are many moment conditions.   Typically, hypothesis tests involving many moment inequalities can have low power, since ``redundant'' moment conditions which are far from binding tend to shift the asymptotic null distribution of the test statistic higher (in a stochastic sense), thus making it harder to reject.  The AS procedure, which combines moment selection (that is, eliminating redundant moment conditions which are far from binding), with hypothesis testing, increases power and yields uniformly asymptotically critical values.

From the above description, we see that the AS procedure
is to evaluate the asymptotic distribution of the test statistic under
a sequence of parameters under the null hypothesis which resemble the sample moment
inequalities, and are drifting to zero.   By doing this, moment
inequalities which are far from binding in the sample (i.e. the
elements of $\mathbf{\hat\mu}$ which are $>>0$) will not contribute to
the asymptotic null distribution of the test statistic, leading to a
(stochastically) smaller distribution and hence smaller critical
values.\footnote{
In contrast, other inequality based testing procedure
(e.g., \citet{Wolak1989}) evaluate the asymptotic distribution of the
test statistic under the ``least-favorable'' null hypothesis
$\mathbf{\mu}=0$, which may lead to very large critical values and low
power against alternative hypotheses of interest.
}

Using the AS procedure requires the specifying an appropriate test statistic for the moment inequalities, and also specifying a drifting sequence of null hypotheses converging to zero.   In doing both we follow the suggestions in AS.   The test statistic in Eq. (\ref{iu_test_stat}) satisfies the requirements for the AS procedure.\footnote{Cf. Eq. (3.6) and Assumption 1 in AS.}  The drafting sequence $\left\{ \kappa_K, K\rightarrow\infty\right\}$ also follows AS's suggestions and is described immediately below.

To obtain valid critical values for $S$ under $H_0$, we use the
following procedure:

\begin{enumerate}
\item Let $D\equiv Diag^{-1/2}(\hat\Sigma)$ denote the diagonal matrix
  with elements $1/\hat\sigma_1,\ldots, 1/\hat\sigma_P$.  Compute $\Omega\equiv D\cdot \hat\Sigma\cdot D$.
\item Compute the vector $\mathbf{\xi}=\kappa_K^{-1}\cdot D\cdot\mathbf{\hat\mu}$ which is equal to
$
\frac{1}{\kappa_{K}}\cdot\left[ \frac{\hat\mu^1}{\hat\sigma_1}, \frac{\hat\mu^2}{\hat\sigma_2}, \cdots , \frac{\hat\mu^P}{\hat\sigma_P} \right]'
$
where $\kappa_K=(\log K)^{1/2}$.\footnote{\citet{AndrewsSoares10} mention several alternative choices for $\kappa_K$. We investigate their performance in the Monte Carlo simulations reported below.} Here $K=N/M$ is the number of trials in each of the $M$ games, and $N$ is the sample size.
\item For $r=1,\ldots, R$, we generate $Z_r\sim N(0, \Omega)$ and compute
$s_r\equiv S(Z_r+ [\mathbf{\xi}]_{+}, \Omega)$, where $[x]_{+}=\max(x,0)$.
\item Take the critical value $c_{1-\alpha}$ as the $(1-\alpha)$-th quantile among $\left\{ s_1, s_2,\ldots, s_R\right\}$.
\end{enumerate}
Essentially, the asymptotic distribution of $S$ is evaluated at the
null hypothesis $[\mathbf{\xi}]_{+}\geq 0$ which, because of the normalizing
sequence $\kappa_K$, is drifting towards zero. In finite samples,
this will tend to increase the number of rejections relative to evaluating the asymptotic distribution at the zero vector.  This is evident
in our simulations, which we turn to next, after describing the test set of games.

\textbf{Interpretation of test results.}
As discussed in the remarks after Proposition \ref{prop_CM_UM}, our test, in essence, checks for violations of the cyclic monotonicity inequalities.
If such violations are substantial, at least one of the players must violate ``incentive compatibility'' conditions on her choices in at least one game cycle, making the joint behavior non-rationalizable by \emph{any} structural QRE satisfying our assumptions. This is much more general than a simple failure to fit a logit QRE.
Moreover, by estimating the left hand sides $\hat{\mu}$ of CM inequalities from the data, we can pin down the exact game cycles that violate CM.

If on the other hand, our test does not reject the QRE hypothesis, our conclusions are less crisp.
Similarly to the results from the revealed preference literature, consistency with CM conditions indicates a possibility result, i.e., in our case, the possibility of existence of a QRE rationalizing the data, rather than yielding estimates of a particular quantal response function or error shock distribution.

\subsection{Test set of two-player games: ``Joker'' games}
As test games, we used a series of four card-matching games where each player has
three choices.  These games are so-called ``Joker'' games which have
been studied in the previous experimental literature
(cf. \citet{ONeill1987} and \citet{BrownRosenthal1990}), and can be considered generalizations of the familiar ``matching pennies'' game in which each player calls out one of three
possible cards, and the payoffs depend on whether the called-out cards
match or not.   Since these games will also form the basis for our laboratory
experiments below, we will describe them in some detail here.\footnote{In our choice of games, we wanted to use simple games comparable to games from the previous literature. Moreover, we wanted our test to be sufficiently powerful. This last consideration steered us away from games for which we know that some structural QRE (in particular, the Logit QRE) performs very well so that the chance to fail the CM conditions is pretty low.}  Table \ref{Table:joker} shows the payoff matrices of the four games which we used in our simulations and experiments.

\begin{table}[htbp]\centering\caption{Four $3\times 3$ games inspired by the Joker Game of \citet{ONeill1987}.}\label{Table:joker}
\begin{center} \scalebox{0.85}{
\begin{tabular}{llcccc}\hline\hline
Game $1$ (Symmetric Joker)& &&  1 & 2 & J \\
&&& [$\bm{1/3}$]\ (.325) & [$\bm{1/3}$] \ (.308) & [$\bm{1/3}$] \ (.367) \\ \hline
 & 1 & [$\bm{1/3}$]\ (.273)  & 10, 30 & 30, 10 & 10, 30 \\
& 2 & [$\bm{1/3}$] \ (.349) & 30, 10 & 10, 30 & 10, 30 \\
& J & [$\bm{1/3}$] \ (.378) & 10, 30 & 10, 30 & 30, 10 \\  \hline
Game $2$ (Low Joker)& &&  1 & 2 & J \\
&&& [$\bm{9/22}$]\ (.359) & [$\bm{9/22}$] \ (.439) & [$\bm{4/22}$] \ (.202) \\ \hline
 & 1 & [$\bm{1/3}$]\ (.253) & 10, 30 & 30, 10 & 10, 30 \\
& 2 & [$\bm{1/3}$] \ (.304)  & 30, 10 & 10, 30 & 10, 30 \\
& J & [$\bm{1/3}$] \ (.442)  & 10, 30 & 10, 30 & 55, 10 \\  \hline
Game $3$ (High Joker)& &&  1 & 2 & J \\
&&& [$\bm{4/15}$]\ (.258) & [$\bm{4/15}$] \ (.323) & [$\bm{7/15}$] \ (.419) \\ \hline
 & 1 & [$\bm{1/3}$]\ (.340) & 25, 30 & 30, 10 & 10, 30 \\
& 2 & [$\bm{1/3}$] \ (.464) & 30, 10 & 25, 30 & 10, 30 \\
& J & [$\bm{1/3}$] \ (.196)  & 10, 30 & 10, 30 & 30, 10 \\  \hline
Game $4$ (Low 2)& &&  1 & 2 & J \\
&&& [$\bm{2/5}$]\ (.487)  & [$\bm{1/5}$] \ (.147) & [$\bm{2/5}$] \ (.366) \\ \hline
 & 1 & [$\bm{1/3}$] \ (.473) & 20, 30 & 30, 10 & 10, 30 \\
& 2 & [$\bm{1/3}$] \ (.220) & 30, 10 & 10, 30 & 10, 30 \\
& J & [$\bm{1/3}$] \ (.307) & 10, 30 & 10, 30 & 30, 10 \\  \hline
\hline
\multicolumn{6}{p{\textwidth}}{\footnotesize \emph{Notes.}
For each game, the unique Nash equilibrium choice probabilities are given in bold font within brackets, while the probabilities in regular font within parentheses are aggregate choice probabilities from our experimental data, described in Section \ref{subsec:exp_design}.} \end{tabular}}
\end{center}
\end{table}

Each of these games has a unique mixed-strategy Nash equilibrium, the
probabilities of which are given in bold font in the margins of the
payoff matrices. Note that the four games in Table \ref{Table:joker} differ only by Row player's payoff.
Nash equilibrium logic, hence, dictates that the Row player's equilibrium
choice probabilities never change across the four games, but that the
Column player should change her mixtures to maintain the Row's
indifference amongst choices.


\textbf{ Joker games have unique regular QRE.}
An important advantage of using Games 1--4 for our application is that in each of our games there is a unique QRE for \emph{any} regular quantal response function (see Appendix \ref{proof:prop:qreUniq} for details).  Regular QRE is an extremely important class of QRE, so let us briefly describe the additional restrictions regularity imposes on the admissible quantal response functions.\footnote{See Appendix \ref{proof:prop:qreUniq} for formal definitions and additional details.} In this paper we focus on testing QRE via its implication of cyclic monotonicity, which involves checking testable restrictions on QRE probabilities when the shock distribution is fixed in a series of games that only differ in the payoffs. These are comparisons \emph{across games}. In the QRE literature, there are typically additional restrictions imposed on quantal response functions \emph{within a fixed game}. In particular, the quantal response functions studied in \citet{GoereeHoltPalfrey05}, in addition to the assumptions we impose in Section \ref{background}, also satisfy the \emph{rank-order property}\footnote{We borrow the name of this property from \citet{Fox07}. \citet{GoereeHoltPalfrey05} call this a ``monotonicity'' property, but we chose not to use that name here to avoid confusion with ``cyclic monotonicity'', which is altogether different.}, which states that actions with higher expected payoffs are played with higher probability than actions with lower expected payoffs. Formally, a quantal response function $\pi_i:\mathbb{R}^{J_{i}}\to\Delta(S_i)$ satisfies the \emph{rank-order property} if for all $i\in N, j,k\in \{1,\ldots,J_i\}$:
\begin{equation}\label{rankorder}
u_{ij}(\mathbf{p})>u_{ik}(\mathbf{p})\Rightarrow \pi_{ij}(\mathbf{p})>\pi_{ik}(\mathbf{p}).
\end{equation}
We stress here that while the rank-order property is not assumed for our test, it is nevertheless a reasonable assumption for QRE, and the inequalities (\ref{rankorder}) can be tested using the same formal statistical framework we described in the previous sections. See Appendix \ref{proof:prop:qreUniq} for details.


{\bfseries Monte Carlo simulations.}
For the Monte Carlo simulations, we first considered artificial data generated under the QRE hypothesis (specifically, under a logit QRE model).
 Table \ref{Table:as_power1} shows the results of the GMS test procedure applied to our setup in terms of the number of rejections. From Table~\ref{Table:as_power1}, we see that the test tends to
(slightly) underreject under the QRE null for most values of the tuning parameter $\kappa_K$.
The results appear relatively robust to changes in $\kappa_K$; a reasonable choice appears to be $\kappa_K=5(\log(K))^{\frac{1}{4}}$, which we will use in our experimental results below.
In a second set of simulations, we generated artificial data under a non-QRE play (specifically, we generated a set of choice probabilities that generate violations of all of the CM inequalities for both players).
The results here are quite stark: in all our simulations, and for all the tuning parameters that we checked, we find that the QRE hypothesis is rejected in every single replication.  Thus our proposed test appears to have very good power properties.

\begin{table}[htbp]\centering\caption{Monte Carlo simulation results under QRE-consistent data}\label{Table:as_power1}
\begin{center} \scalebox{0.85}{
\begin{tabular}{cccccc}\hline\hline
         &     & \multicolumn{4}{c}{Tuning parameter $\kappa_K$} \\ \cline{3-6}
$N$ & \# rejected${}^{a}$ at & $5(\log(K))^{\frac{1}{2}}$ & $5(\log(K))^{\frac{1}{4}}$ & $5(\log(K))^{\frac{1}{8}}$ & $5(2\log \log(K))^{\frac{1}{2}}$ \\\hline
1000              & 5\% & 7 & 9  & 13  & 7  \\
                  & 10\% & 13 & 17 & 26 & 14 \\
                  & 20\% & 33 & 54 & 68 & 46 \\ \hline
5000              & 5\% &13  &32  &43   & 21 \\
                  & 10\% &26  &55  & 77 & 41  \\
                  & 20\% &61  & 102 & 124 & 85 \\ \hline
9000              & 5\% & 18 & 40 & 51 & 32  \\
                 & 10\% & 33 & 61 & 79 & 53 \\
                 & 20\% & 65 & 114 & 143 & 97 \\
                 \hline\hline
\multicolumn{6}{p{0.8\textwidth}}{\footnotesize \emph{Notes.}
    $K=\frac{N}{4}$ is the total number of rounds of each of the four
    games. All numbers in columns 3--6 are observed rejections out of
    $500$ replications. All computations use $R=1000$ to simulate the
    corresponding critical values. \newline
    ${}^{a}$: \# rejected out of 500 replications for each significance level.}
\end{tabular}}
\end{center}
\end{table}

\begin{figure}[htb]
\centering
\scalebox{0.9}{
\includegraphics{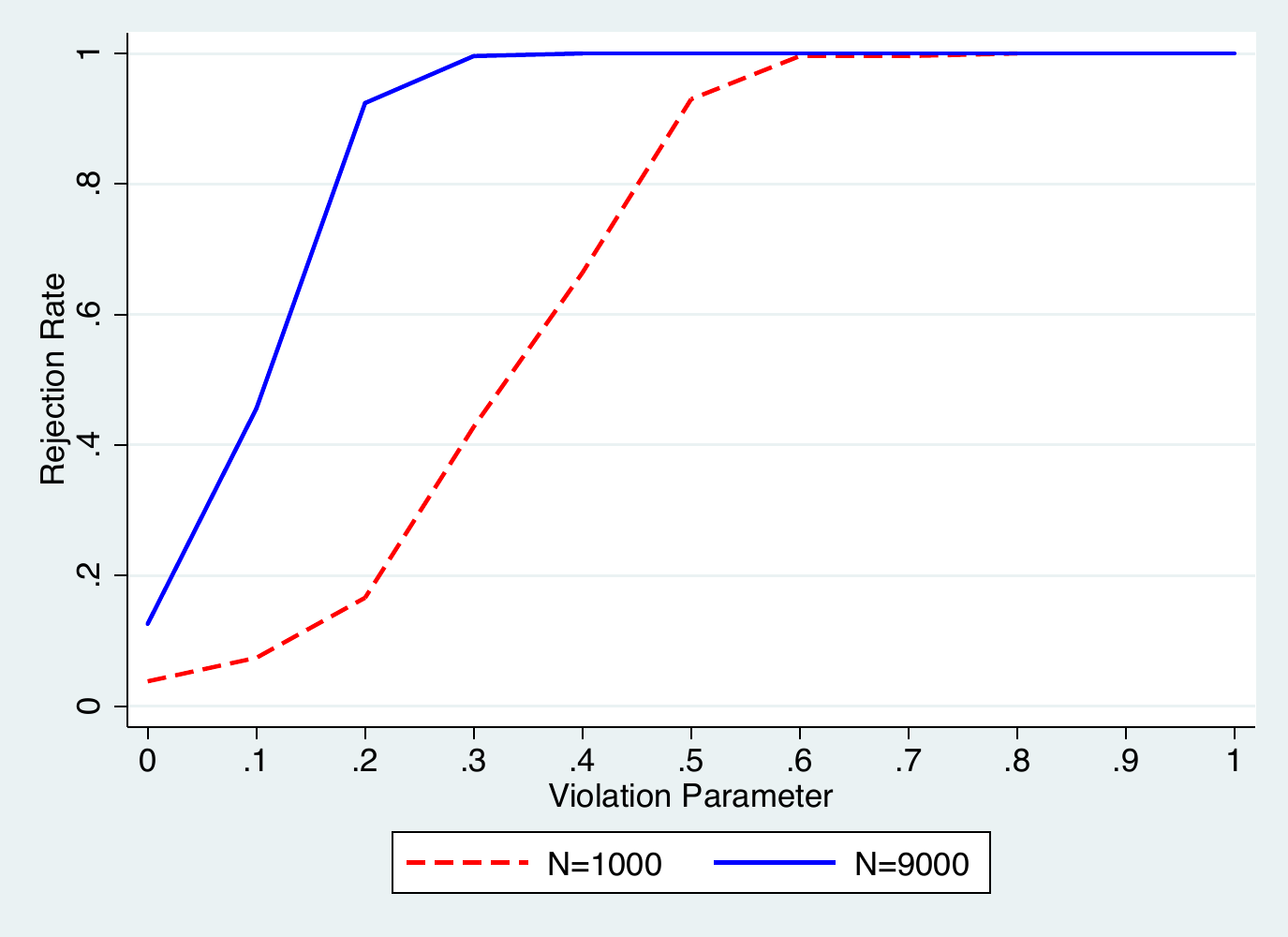}}
\caption{Rejection rate at 10\% level as a function of violation parameter $\lambda$ in Monte Carlo simulations for different $N$ ($\kappa_K=5(\log(K))^{\frac{1}{4}}$.) } \label{fig:power_graph}
\end{figure}

To illustrate the power properties further, we constructed ``mixed'' datasets consisting of linear combinations of the two sets of choice probabilities, one QRE-consistent, and one QRE-inconsistent (i.e., the one that generates violations of all CM inequalities and the one that results in all CM inequalities satisfied) and varied the relative weight, $\lambda$, from 0 to 1. 
Thus for $\lambda=0$ we have a fully QRE-consistent distribution, for $\lambda=1$ we have a distribution completely inconsistent with QRE, with $0<\lambda<1$ spanning ``mixed'' distribution in which some cyclic monotonicity inequalities are violated, and some are not. The corresponding rejection probabilities are graphed in Figure \ref{fig:power_graph}, and show that our test appears keenly sensitive to violations of QRE; rejection probabilities are high once the mixing parameter $\lambda$ exceeds 0.2 (for large $N=9000$, and 0.4 for small $N=1000$), and for values of $\lambda$ exceeding 0.3 (0.6 for small $N$), the test always rejects.   This demonstrates the good power properties of our testing procedure.

\section{Experimental evidence}\label{exp_design}

In this section we describe an empirical application of our test to data generated from laboratory experiments.   Lab experiments appear ideal for our test because the invariance of the distribution of utility shocks across games (Assumption \ref{assumption:invariance}) may be more likely to hold in a controlled lab setting than in the field. This consideration also prevented us from using the experimental data from published $3\times3$ games, as those usually do not have the same subjects participating in several (we require up to 4) different games in one session.


Our testing procedure can be applied to the experimental data from Games 1--4 as follows.
As defined previously, let the $P$-dimensional vector $\bm\nu$ contain the value of the CM inequalities evaluated at the choice frequencies observed in the experimental data. Using our four games, we can construct cycles of length $2, 3$, and $4$. Thus we have $12$ possible orderings of $2$-cycles, $24$ possible orderings of $3$-cycles, and $24$ possible orderings of $4$-cycles. Since CM inequalities are invariant to the change of the starting game index, it is sufficient to consider the following $20$ cycles of Games $1-4$:
\begin{IEEEeqnarray}{lCl}
121, 131, 141, 232, 242, 343 \nonumber \\
1231, 1241, 1321, 1341, 1421, 1431, 2342, 2432 \nonumber \\
12341, 12431, 13241, 13421, 14231, 14321 \label{ordered_cycles}
\end{IEEEeqnarray}
Moreover, these 20 cycles are distinct depending on whether we are considering the actions facing the Row or Column player (which involve different payoffs); thus the total number of cycles across the four games and the two player roles is $P=40$.  This is the number of coordinates of vector $\bm\nu$, defined in \eqref{cm_test_explicit}. Additional details on the implementation of the test, including
explicit expressions for the variance-covariance matrix of $\nu$, are
provided in Appendix \ref{appxC}.


\subsection{Experimental design}\label{subsec:exp_design}

The subjects in our laboratory experiments were undergraduate students at
the University of California, Irvine, and all experiments were conducted
at the ESSL lab there. We have conducted a total of 3 
sessions where in each session subjects played one the following sequences of games from  Table \ref{Table:joker}: 12, 23, and 3412.
In the first two sessions, subjects played 20 rounds per game; in the last session subjects played 10 rounds per game.\footnote{We had to adjust the number of rounds because of the timing constraints.}   Across all three sessions, there was a total of 96 subjects.
To reduce repeated game effects, subjects were randomly rematched each round.
To reduce framing effects, the payoffs for every subject were displayed as payoffs for the Row player, and actions were abstractly labelled $A$, $B$, and $C$ for the Row player, and $D$, $E$, and $F$ for the Column player.

In addition to recording the actual choice frequencies in each round of
the game, we periodically also asked the subjects to report their beliefs
regarding the likelihood of their current opponent playing each of the
three strategies. Each subject was asked this question once s/he had
chosen her action but before the results of the game were displayed. To
simplify exposition, we used a two-thumb slider which allowed subjects to
easily adjust the probability distribution among three choices. Thus we
were able to compare the CM tests based on subjective probability
estimates with the ones based on actual choices.\footnote{We chose not to
  incentivize belief elicitation rounds largely to avoid imposing extra
  complexity on the subjects.
Thus our results using elicited belief estimates should be taken with some
caution. On the other hand, if what we elicited was completely meaningless, we would not observe as much QRE consistency as we do in our subject-by-subject results below.}

Subjects were paid the total sum of payoffs from all rounds, exchanged into U.S. dollars using the exchange rate of 90 cents for 100 experimental currency units, as well as a show-up fee of \$7.
The complete instructions of the experiment are provided in Appendix \ref{appx}.

\subsection{Results}\label{exp_results}

We start analyzing the experimental data by reporting the aggregate choice frequencies in Games $1$--$4$ in Table \ref{Table:joker} alongside Nash equilibrium predictions. Comparing theory with the data, we see that there are a lot of deviations from Nash for both Column and Row players.

Table \ref{Table:as_estimates} is our main results table. It shows the test results of checking the cyclic monotonicity conditions with our experimental data.

\begin{table}[htbp]\centering\caption{Testing for Cyclic Monotonicity in Experimental Data: Generalized
    Moment Selection}\label{Table:as_estimates}
\begin{center} \scalebox{0.75}{
\begin{tabular}{crccccc}\hline\hline
{\bfseries Data sample}   & {\bfseries $AS$ test stat} & {\bfseries $c_{0.95}^R$} & &&& \\ \hline
{\underline{\bfseries All subjects pooled:}} \\
All cycles &  68.194 & 29.985 &&&& \\
Row cycles &  68.194 & 26.265 &&&& \\
Col cycles &  0.000 & 6.835 &&&& \\ \hline\hline
&&&&&&\\
{\underline{\bfseries Subject-by-subject:}}  &  {\bfseries Avg $AS$} & {\bfseries Avg $c_{0.95}^R$} & \multicolumn{3}{c}{\bfseries \# rejected} & {\bfseries Avg CM violations}\\
  & & & at 5\% & at 10\% & at 20\%  & (\% of total) \\ \hline
{{\bfseries Subj. v. self$^a$}}  \\
(Total subj.: $96$) \\
All cycles &  212.570 & 18.538 & 29 & 37 & 41 & 41.59 \\
Row cycles &  203.108 & 12.421 & 20 & 26 & 31 & 44.53\\
Column cycles & 9.462 & 11.849 & 18 & 19 & 21 & 38.65 \\ \hline
{{\bfseries Subj. v. others$^b$}}   \\
(Total subj.: $96$)  \\
Row cycles                      & 3.936  & 10.920 & 7 & 11  & 15 & 35.00 \\
(Total subj.: $96$)  \\
Col cycles  & 103.872  & 11.734 & 16 & 18 & 22 & 38.70 \\
\hline
{{\bfseries Subj. v. beliefs$^c$}}  \\
(Total subj.: $59$)  \\
Row cycles & 3.681 & 6.883 & 5 & 5 & 5 & 33.051 \\
(Total subj.: $61$) \\
Col cycles & 9.622 & 8.732 & 10 & 13 & 14 & 35.000 \\ \hline\hline
\multicolumn{7}{p{1.3\textwidth}}{{\footnotesize \emph{Notes.} All computations use $R=1,000$. In subject-by-subject computations some subjects in some roles exhibited zero choice variance, so in those cases we replaced the corresponding (ill-defined) elements of $Diag^{-1/2}(\hat\Sigma)$ with ones and when computing the test statistic, left out the corresponding components of $\hat{\mu}$.
The tuning parameter in AS procedure was set equal to $\kappa_z=5(\log(z))^{\frac{1}{4}}$. \newline
\text{$^{a}$}v. self: the opponent's choice frequencies are obtained from the same subject playing the respective opponent's role. \newline
\text{$^{b}$}v. others: the opponent's choice frequencies are averages over the subject's actual opponents' choices when the subject was playing her respective role. 
\newline
\text{$^{c}$}v. beliefs: the opponent's choice frequencies are averages over the subject's elicited beliefs about the opponent choices when the subject was playing the respective role (since belief elicitation rounds were fixed at the session level, subjects' beliefs may not be elicited in some roles and some games. We dropped them from the analysis).
 }}
\end{tabular}}
\end{center}
\end{table}

Based on our current dataset, we find that QRE is soundly rejected for the pooled data (with test statistic 68.194 and 5\% critical value 29.985). This may not be too surprising, since in our design subjects experience both player roles (Row and Column), and so this pooled test imposes the auxiliary assumption on all subjects being homogeneous across roles in that their utility shocks are drawn from identical distributions.

Therefore, in the remaining portion of Table \ref{Table:as_estimates}, we test the QRE hypothesis separately for different subsamples of the data.  First, we consider separately the CM inequalities pertaining to Row players and those pertaining to Column players.\footnote{Note that the sum of the test statistics corresponding to the Column and Row inequalities sum up to the overall test statistic; this is because the Row and Column inequalities are just subsets of the full set of inequalities.}
By doing this, we allow the utility shock distributions to differ depending on a subject's role (but conditional on role, to still be identical across subjects).

We find that while we still reject QRE\footnote{Strictly speaking, this is no longer a test of QRE, because by restricting attention to cycles pertaining to only one player role, we essentially consider only one-player equilibrium version of QRE, which is more akin to a discrete choice problem.} for the Row players, we cannot do so for Column players. Thus overall QRE-inconsistency is largely due to the behavior of the Row players.
Seeing that Row players' inequalities are violated more often than Column players' inequalities suggests that Row players' choice probabilities do not always adjust toward higher-payoff strategies.
That the violations come predominantly from the choices of Row players is interesting because, as we discussed
above, the payoffs are the same across all the games in
our experiment for the Column player, but vary across games for the Row player.

Some intuition for this may come from considering the nature of the Nash equilibria in these games.
The (unique) mixed strategy Nash equilibrium prescribes mixing probabilities which are the same across games for the Row player, but vary across games for the  Column player -- since in equilibrium each player's mixed strategy probabilities are chosen so as to make the {\em opponent} indifferent between their actions (cf. \citet{GoereeHolt01}).  This is clear from Table \ref{Table:joker}, where the Nash equilibrium probabilities are given.
The greater degree of violations observed for the Row play may reflect an intrinsic misapprehension of this somewhat paradoxical logic of Nash equilibrium play, and sensitivity to change in own payoffs that carries over from Nash to noisier quantal response equilibria.\footnote{
From a theoretical point of view, this observation is also consistent with \citet{Golman11} who showed that in heterogeneous population games a representative agent for a pool of individuals may not be described by a structural quantal response model even if all individuals use quantal response functions (in the sense that with at least four pure strategies there is no iid payoff shock structure that generates the representative quantal response function). 
Games with three pure strategies like in our experiment usually fail to have a representative agent, too, indicating a need to take into account heterogeneity of the player roles.
}

Continuing in this vein, the lower panel of Table \ref{Table:as_estimates} considers tests of the QRE hypothesis for each subject individually.  Obviously, this allows the distributions of the utility shocks to differ across subjects. For these subject-by-subject tests, there is a question about how to determine a given's subject beliefs about her opponents' play. We consider three alternatives: (i) set beliefs about opponents equal to the subject's own play in the opponent's role; (ii) set beliefs about opponents equal to opponents' actual play (i.e., as if the subject was playing against an average opponent); and (iii) set beliefs about opponents equal to the subject's elicited beliefs regarding the opponents' play.

The results appear largely robust across these three alternative ways of accounting for subjects' beliefs.  We see that we are not able to reject the QRE hypothesis for most of the subjects, for significance levels going from 5\% to 20\%.   When we further break down each subject's observations depending on his/her role (as Column or Row player), thus allowing the utility shock distributions to differ not only across subjects but also for each subject in each role, the number of rejections decreases even more.  Curiously, we see that in the subject vs. self results, the Row inequalities generate more violations, while the Column inequalities generate more violations in the subject vs. others results. This pattern is consistent with our findings for the pooled data: Since it is the Row players' inequalities that are predominantly violated in the pooled data, these violations only intensify in subject vs. self for the Row cycles. In subject vs. others, however, the inequalities are computed under beliefs corresponding to the average opponent's actual behavior in the \emph{opponent role}, e.g., for Row players the average Column behavior and vice versa. The set of all cyclic monotonicity inequalities is then further partitioned into row and column cycles.
Thus for Row players the row cycles reflect the Column behavior, which is more consistent with QRE and so shows less violations than the column cycles for the Column player, which reflect the more erratic Row behavior.

One caveat here is that when we are testing on a subject-by-subject basis, we are, strictly speaking, no longer testing an equilibrium hypothesis, because we are not testing -- and indeed, {\em cannot} test given the randomized pairing of subjects in the experiments -- whether the given subject's opponents are playing optimally according to a QRE.   Hence, our tests should be interpreted as tests of subjects' ``better response'' behavior given beliefs about how their opponents' play.

The general trend of these findings -- that the QRE hypothesis appears more statistically plausible once we allow for sufficient heterogeneity across subjects and across roles -- confirms existing results in \citet{McKelveyPalfreyWeber00} who, within the parametric logit QRE framework and $2\times2$ asymmetric matching pennies, also found evidence increasing for the QRE hypothesis once subject-level heterogeneity was accommodated.

{\bfseries Robustness check: Nonlinear utility and risk aversion.}
Our test results above are computed under the assumption of risk-neutrality. \citet{GoereeHoltPalfrey00} have shown that allowing for nonlinear utility (i.e. risk aversion) greatly improves the fit of QRE to experimental evidence.
Since our test can be applied under quite general specification of payoff functions, to see the effects of risk aversion on the test results we recomputed the test statistics under an alternative assumption that for each player, utility from a payoff of $x$ is $u(x)=x^{1-r}$, where $r\in[0,1)$ is a constant relative risk aversion factor.\footnote{For $r=1$ the log-utility form is used. In our computations, we restrict the largest value of $r$ to $0.99$ to avoid dealing with this issue.} %
Here, we computed the test statistics and critical values for values of $r$ ranging from
$0$ to $0.99$.

When we pool all the subjects together, we find results very similar to what is reported in Table \ref{Table:as_estimates}: QRE is rejected when all cycles are considered; it is also rejected when only the Row cycles are considered; it cannot be rejected when only the Column cycles are considered, for all values of $r\in[0,0.99]$. Thus we do not observe any risk effects in the pooled data.\footnote{ For space reasons, we have not reported all the test statistics and critical values, but they are available from the authors upon request.}

\begin{figure}[htb]
\centering
\scalebox{0.9}{
\includegraphics{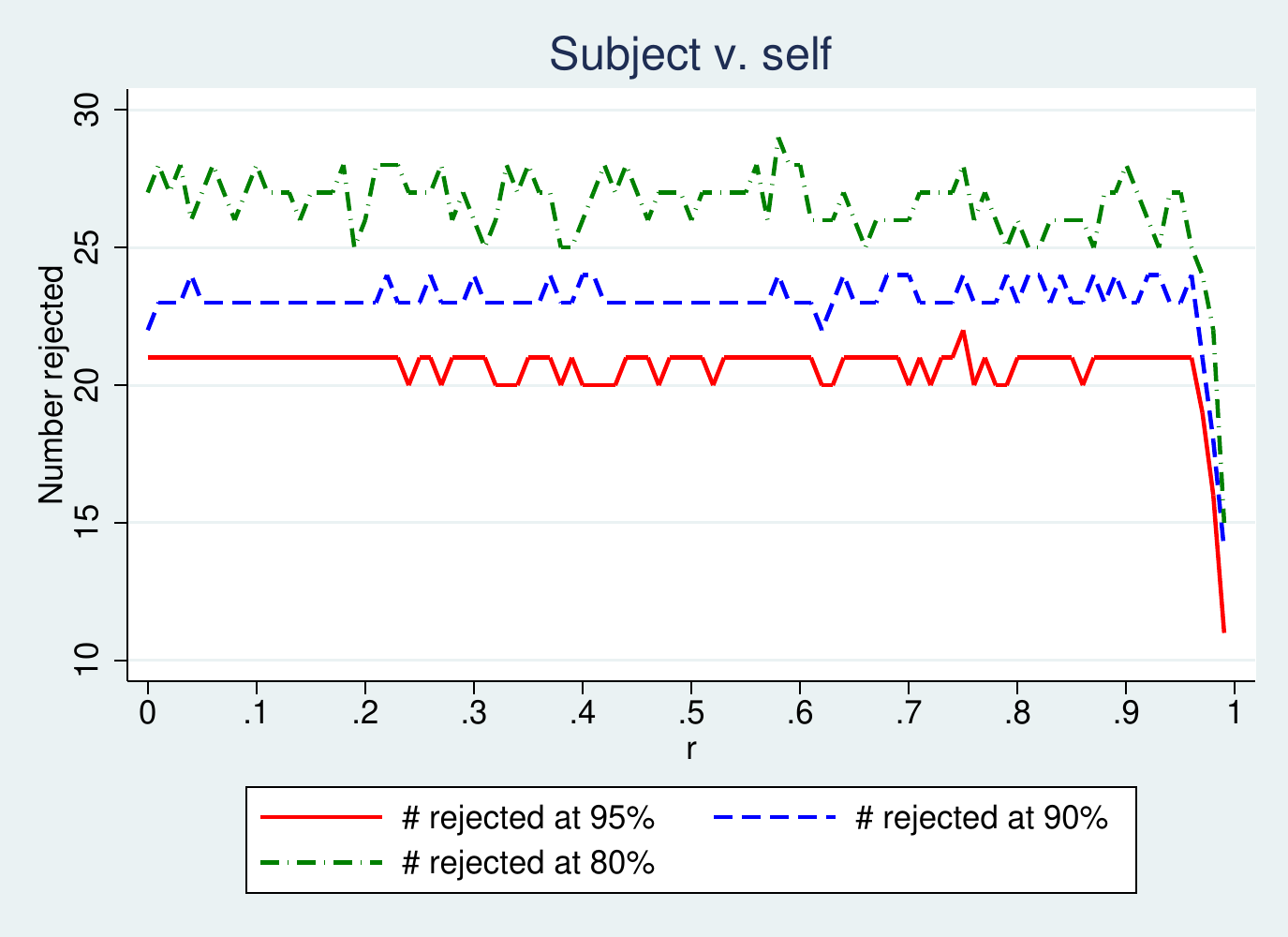}}
\caption{Effects of risk aversion on subject-by-subject rejections} \label{fig:rejections_by_risk}
\end{figure}

Breaking down these data on a subject-by-subject basis, we once again see that allowing for risk aversion does not change our previous results obtained under the assumption of linear utility. Specifically, as graphed in Figure \ref{fig:rejections_by_risk}, the number of rejections of the QRE hypothesis for the ``subject vs. self'' specification is relatively stable for all $r<0.99$, staying at about 21 rejections at $5\%$ level,  at about 23 rejections at $10\%$ level, and between 25 and 30 rejections at $20\%$ level.
Thus our analysis here suggests the our test results are not driven by risk aversion.

This robustness to risk aversion might seem surprising at first (e.g., the analysis in \citet{GoereeHoltPalfrey03} shows that risk aversion might be an important factor in fitting logit QRE in several bimatrix games). 
However, introducing constant relative risk aversion has limited effects on the cyclic monotonicity inequalities as it only changes the scale of payoffs, with little effects on the relative difference between payoffs across games in a cycle, whenever the relative differences are not large to begin with, as in our games.
Similarly, our test results depend less on the exact form of the utility function as long as differences in payoffs across the games are not too dramatic.

Of course, risk aversion might be an important factor in other games, so checking for the potential effects of risk aversion on test results might be a necessary post-estimation step.


\section{Conclusions and Extensions}\label{concl_exte}
In this paper we present a new nonparametric approach for testing the QRE hypothesis in finite normal form games.  
We go far beyond consistency with the usual logit QRE by allowing players to use any structural quantal response function satisfying mild regularity conditions.
This flexibility comes at the cost of requiring the payoff shocks to be fixed across a series of games.
The testing approach is based on moment inequalities derived from the {\em cyclic monotonicity} condition, which is in turn derived from the convexity of the random utility model underlying the QRE hypothesis. We investigate the performance of our test using a lab experiment where subjects play a series of generalized matching pennies games.

While we primarily focus on developing a test of the QRE hypothesis in games involving two or more players, our procedure can also be applied to situations of stochastic individual choice.
Thus our test can be viewed more generally as a semiparametric test of quantal response, and, in particular, discrete choice models.
Moreover, in finite action games as considered here, QRE has an
identical structure to Bayesian Nash equilibria in discrete games of incomplete
information which have been considered in the empirical industrial
organization literature (e.g. \citet{Bajarietal2010},
\citet{DePaulaTang2012}, or \citet{LiuVuongXu2013}). Our approach can
potentially be useful for specification testing in those settings;
however, as we remarked above, one hurdle to implementing such tests on
field data is the possibility of multiple equilibria played in the data.  Adapting these
tests to allow for multiple equilibria is a challenging avenue for future research.


\phantomsection
\addcontentsline{toc}{section}{References}
\singlespacing
\setlength{\bibsep}{0.1pt}

\newpage
\appendix\singlespace\small

{\bfseries\large Appendix: supplemental material (for online publication)}

\section{Cyclic Monotonicity and Utility Maximization}\label{CM_UM}

In this section we show  that  the CM inequalities are equivalent to players' utility maximization. In order to establish this result we exploit the fact that the set of  QRE  can be seen as the set of NE of a perturbed game. In particular, it can be shown\footnote{In particular, the papers by \citet[Prop. 3]{CominettiMeloSorin(10)}, \citet[Thm. 1]{HofSan02}, and \citet[Remark 2.2]{MetikopoulosSandholm15} establish a one-to-one correspondence between the Nash equilibria of the game with payoffs (\ref{PerPayoffs}) and QRE. } that QRE corresponds  to the set of NE of a game where players'  payoffs are given by
\begin{equation}\label{PerPayoffs}
\G^i(\bm{\pi}):=\big<\bm{\pi}_i,\bm{u}_i\big>-\tilde{\varphi}^i(\bm{\pi}_i),\quad\forall i\in N,
\end{equation}
\noindent  with $\tilde{\varphi}^i(\bm{\pi}_i)$ corresponding  to the  Fenchel-Legendre conjugate (hereafter convex conjugate) of the function $\varphi^i(\bm{u}_i)$ defined in \eqref{eq:phi}.\footnote{Formally,  for a convex function $f$ the  Fenchel-Legendre conjugate is defined by $\tilde{f}(\bm{y})=\sup_{\bm{x}}\{\big<\bm{y},\bm{x}\big>-f(\bm{x})\}$.}


\noindent\emph{Proof of Proposition \ref{prop_CM_UM}:}
\emph{Utility maximization implies CM}: Consider a cycle of length $\mathcal{L}-1$ with   $[\bm{u}_i]^m$ and $[\bm{\pi}^*_i]^m$ denoting  the expected payoffs and equilibrium probabilities in a game indexed $m$ in the cycle, respectively. By utility maximization, it is easy to see that  for each player $i$  the following inequalities must hold:
 \begin{equation}\label{UMinequalities}
\big<[\bm{u}_i]^{m+1},[\bm{\pi}^*_i]^{m}\big>-\tilde{\varphi}^i([\bm{\pi_i}^*]^m)\leq \big<[\bm{u}_i]^{m+1},[\bm{\pi}^*_i]^{m+1}\big>-\tilde{\varphi}^i([\bm{\pi}^*_i]^{m+1}), \quad \forall m.
\end{equation}
Rewriting as
 \begin{equation*}
\big<[\bm{u}_i]^{m+1}-[\bm{u}_i]^{m},[\bm{\pi}^*_i]^{m}\big> \leq \big<[\bm{u}_i]^{m+1},[\bm{\pi}^*_i]^{m+1}\big>-
\big<[\bm{u}_i]^{m},[\bm{\pi}^*_i]^{m}\big> +\tilde{\varphi}^i([\bm{\pi_i}^*]^m) - \tilde{\varphi}^i([\bm{\pi}^*_i]^{m+1}),
\end{equation*}
and adding up over the cycle, we get the CM inequalities \eqref{basicCM}. \\

\emph{CM implies utility maximization}:  Suppose that CM holds and let $[\bm{u}_i]^m$ denote expected utility in game $m$. Thanks to CM we know that  $[\bm{\pi}^*_i]^m=\nabla \varphi^i([\bm{u}_i]^m)$ for all $m$. Now by  Fenchel's equality 
it follows that $[\bm{\pi}^*_i]^m=\nabla \varphi^i\left([\bm{u}_i]^m\right)$ iff  
$$\big<[\bm{u}_i]^{m},[\bm{\pi}^*_i]^{m}\big>-\tilde{\varphi}^i([\bm{u}_i]^m)=\varphi([\bm{u}_i]^m).$$

The last expression implies that $[\bm{\pi}_i^*]^m\in \arg\max_{\bm{\pi}_i}\{\G^i(\bm{\pi})\}$. Thus we conclude that CM implies Utility maximization. \hfill$\square$


Intuitively, the set of inequalities \eqref{UMinequalities} can be seen as set of incentive compatibility constraints across the series of  games that only differ in the payoffs. This means that our CM conditions capture players' optimization behavior with respect to changes in expected payoffs across such games.

 

\section{Proof of Proposition \ref{prop1}}\label{proof:prop1}

Suppose there are two games that differ only in the payoffs. For $M=2$, the cyclic monotonicity condition \eqref{cm_test} reduces to
\begin{IEEEeqnarray*}{rCl}
\sum_{j=1}^{J_i} (u_{ij}^1- u_{ij}^0)\pi_{ij}^0+ \sum_{j=1}^{J_i} (u_{ij}^0- u_{ij}^1)\pi_{ij}^1 \leq 0
\end{IEEEeqnarray*}
or, equivalently,
\begin{IEEEeqnarray}{rCl}
\sum_{j=1}^{J_i} (u_{ij}^1- u_{ij}^0)(\pi_{ij}^0-\pi_{ij}^1) \leq 0 \label{cond2}
\end{IEEEeqnarray}
Suppose that the RHS of \eqref{cond1} is non-negative. Then HHK condition \eqref{rcTestHHK} implies CM.
To see this, notice that for non-negative utilities differences in \eqref{cond1}
\begin{IEEEeqnarray*}{rCl}
(u_{i1}^1- u_{i1}^0)(\pi_{i1}^0-\pi_{i1}^1) \leq 0
\end{IEEEeqnarray*}
by HHK condition for $k=1$. Then
\begin{IEEEeqnarray*}{rCl}
(u_{i2}^1- u_{i2}^0)(\pi_{i2}^0-\pi_{i2}^1) + (u_{i1}^1- u_{i1}^0)(\pi_{i1}^0-\pi_{i1}^1) &\leq& \\
(u_{i2}^1- u_{i2}^0)(\pi_{i2}^0-\pi_{i2}^1) + (u_{i2}^1- u_{i2}^0)(\pi_{i1}^0-\pi_{i1}^1) &=& \\
(u_{i2}^1- u_{i2}^0)((\pi_{i1}^0+\pi_{i2}^0)-(\pi_{i1}^1+\pi_{i2}^1)) &\leq& 0
\end{IEEEeqnarray*}
where the last inequality follows from HHK condition for $k=2$ and $u_{i2}^1- u_{i2}^0\geq 0$. Repeating the same procedure for $k=3,\ldots, J_i$, we obtain the CM condition \eqref{cond2} for $M=2$.

Conversely, suppose that \eqref{cond2} holds. For the case of two games, \eqref{cond2} holding for all players is necessary and sufficient to generate QRE-consistent choices. All premises are satisfied for HHK's Theorem $2$, so condition \eqref{rcTestHHK} follows.
One can also show it directly. Clearly, given \eqref{cond2}, we can always re-label strategy indices so that \eqref{cond1} holds. Let $k=1$ and by way of contradiction, suppose that \eqref{rcTestHHK} is violated, i.e.
$\pi_{i1}^1-\pi_{i1}^0<0$.  Since \eqref{cond2} holds, the probabilities in both games are generated by a QRE.
Due to indexing in \eqref{cond1},
$$
u_{i1}^1- u_{ij}^1 \geq u_{i1}^0- u_{ij}^0
$$
for all $j>1$. But then by definition of QRE in \eqref{qre_p}, $\pi_{i1}^1\geq \pi_{i1}^0$. Contradiction, so \eqref{rcTestHHK} holds for $k=1$. By induction on the strategy index, one can show that \eqref{rcTestHHK} holds for all $k\in\{1,\ldots,J_i\}$.  This completes the proof. \hfill$\square$

\section{Uniqueness of Regular QRE in Experimental Joker Games}\label{proof:prop:qreUniq}

In this section we show formally that any regular QRE in Games 1--4 from Table \ref{Table:joker} is unique.
We start by recalling the necessary definitions.

A quantal response function $\pi_i:\mathbb{R}^{J_{i}}\to\Delta(S_i)$ is \emph{regular}, if it satisfies Interiority, Continuity, Responsiveness, and Rank-order\footnote{This property is called Monotonicity in \citet{GoereeHoltPalfrey05}.} axioms \citep[p.355]{GoereeHoltPalfrey05}.
Interiority, Continuity, and Responsiveness are satisfied automatically under the \emph{structural approach} to quantal response\footnote{In this approach, the quantal response functions are derived from the primitives of the model with additive payoff shocks, as described in Section \ref{background}.} that we pursue in this paper as long as the shock distributions have full support.
Importantly, for some shock distributions this approach may fail to satisfy the Rank-order Axiom, i.e. the intuitive property of QRE saying that actions with higher expected payoffs are played with higher probability than actions with lower expected payoffs. For the sake of convenience, we repeat the axiom here.

A quantal response function $\pi_i:\mathbb{R}^{J_{i}}\to\Delta(S_i)$ satisfies \emph{Rank-order Axiom} if for all $i\in N, j,k\in \{1,\ldots,J_i\}$ $u_{ij}(\mathbf{p})>u_{ik}(\mathbf{p})\Rightarrow \pi_{ij}(\mathbf{p})>\pi_{ik}(\mathbf{p})$.

Notice that the Rank-order Axiom involves comparisons of expected payoffs from choosing different pure strategies \emph{within} a fixed game. As briefly discussed in Section \ref{sec:test_limitations}, consistency of the data with the Rank-order Axiom can be tested: the Axiom is equivalent to the following inequality for each player $i$ and pair of $i$'s strategies $j,k \in \{1,\ldots,J_i\}$:
\begin{equation}
(u_{ij}(\mathbf{p})-u_{ik}(\mathbf{p}))(\pi_{ij}(\mathbf{p})-\pi_{ik}(\mathbf{p}))\geq 0 \label{eq:monotonicityTest}
\end{equation}
Thus a modified test for consistency with a \emph{regular} QRE involves two stages: first, check if the data are consistent with a structural QRE using the cyclic monotonicity inequalities (which compare choices across games) as described in Section \ref{stat_properties}, and second, if the test does not reject the null hypothesis of consistency, check if the Rank-order Axiom (which compares choices within a game) holds by estimating \eqref{eq:monotonicityTest} for each game.\footnote{The test procedure is similar to the one in Section \ref{stat_properties}, with an appropriately modified Jacobian matrix.}

Alternatively, the Rank-order Axiom can be imposed from the outset by making an extra assumption about the shock distributions. In particular, \citet[Proposition 5]{GoereeHoltPalfrey05} shows that under the additional assumption of exchangeability, the quantal response functions derived under the structural approach are regular.

Notice that Assumption \ref{assumption:invariance} (Invariance) is not required for the test of the Rank-order Axiom. In theory, we may have cases where the data can be rationalized by a structural QRE that fails Rank-order, by a structural QRE that satisfies it (i.e., by a regular QRE), or by quantal response functions that satisfy Rank-order in each game but violate the assumption of fixed shock distributions across games. In the latter case, checking consistency with other boundedly rational models (e.g., Level-$k$ or Cognitive Hierarchy) becomes a natural follow-up step.

We can now turn to the uniqueness of the regular QRE in our test games.

\begin{thm}\label{prop:qreUniq}
 For each player $i\in N\equiv \{\text{Row}, \text{Col}\}$ fix a regular quantal response function $\pi_i:\mathbb{R}^{J_i}\to \Delta(S_i)$ and let $Q\equiv (\pi_i)_{i\in N}$. In each of Games 1--4 from Table \ref{Table:joker} there is a unique quantal response equilibrium $(\sigma^R,\sigma^C)$ with respect to $Q$.
 Moreover, in Game 1, the unique quantal response equilibrium is $\sigma^R=\sigma^C=\left(\frac{1}{3},\frac{1}{3},\frac{1}{3}\right)$.
 In Game 2, $\sigma^R_1=\sigma^R_2\in\left(0,\frac{1}{3}\right)$ and $\sigma^C_1=\sigma^C_2\in\left(\frac{1}{3},\frac{9}{22}\right]$.
 In Game 3, $\sigma^R_1=\sigma^R_2\in\left(\frac{1}{3}, 1\right)$ and $\sigma^C_1=\sigma^C_2\in\left[\frac{4}{15},\frac{1}{3}\right)$.
 In Game 4, $\sigma^R_2=\sigma^R_J\in\left(0,\frac{1}{3}\right)$ and $\sigma^C_1=\sigma^C_J\in\left(\frac{1}{3},\frac{2}{5}\right]$.
\end{thm}

Notice that the equilibrium probability constraints in Proposition \ref{prop:qreUniq} hold in any regular QRE, not only logit QRE. For the logit QRE they hold for any scale parameter $\lambda\in[0,\infty)$.

\begin{proof}
In order to prove uniqueness and bounds on QRE probabilities we will be mainly using Rank-order and Responsiveness properties of a regular QRE.

Suppose Row plays $\mathbf{\sigma}^R=(\sigma^R_1, \sigma^R_2, \sigma^R_J)$, Col plays $\mathbf{\sigma}^C=(\sigma^C_1, \sigma^C_2, \sigma^C_J)$, and $(\mathbf{\sigma}^R, \mathbf{\sigma}^C)$ is a regular QRE.\footnote{Obviously, $\sigma^R_J=1-\sigma^R_1-\sigma^R_2$ and $\sigma^C_J=1-\sigma^C_1-\sigma^C_2$.}
Expected utility of Col from choosing each of her three pure strategies in any of Games 1--4 (see payoffs in Table \ref{Table:joker}) is
\begin{IEEEeqnarray*}{rCl}
  u_{C1}(\mathbf{\sigma}^R)&=& 30-20\sigma^R_2 \\
  u_{C2}(\mathbf{\sigma}^R)&=& 30-20\sigma^R_1 \\
  u_{CJ}(\mathbf{\sigma}^R)&=& 10+20\sigma^R_1+20\sigma^R_2
\end{IEEEeqnarray*}
Consider Game 1. Expected utility of Row in this game from choosing each of her three pure strategies is
\begin{IEEEeqnarray*}{rCl}
  u_{R1}(\mathbf{\sigma}^C)&=& 10+20\sigma^C_2 \\
  u_{R2}(\mathbf{\sigma}^C)&=& 10+20\sigma^C_1 \\
  u_{RJ}(\mathbf{\sigma}^C)&=& 30-20\sigma^C_1-20\sigma^C_2
\end{IEEEeqnarray*}
Consider Row's equilibrium strategy. There are two possibilities: 1) $\sigma^R_1> \sigma^R_2$. Then Rank-order applied to Col implies $\sigma^C_2< \sigma^C_1$. Now Rank-order applied to Row implies
    $\sigma^R_1< \sigma^R_2$. Contradiction. 2) $\sigma^R_1< \sigma^R_2$. Then Rank-order applied to Col implies
    $\sigma^C_2> \sigma^C_1$. Now Rank-order applied to Row implies     $\sigma^R_1> \sigma^R_2$. Contradiction.
Therefore, in \emph{any} regular QRE in Game 1, $\sigma^R_1=\sigma^R_2$, and consequently, $\sigma^C_1=\sigma^C_2$.
Suppose $\sigma^R_1>\frac{1}{3}$. Then Rank-order applied to Col implies $\sigma^C_J > \sigma^C_1$, and since  $\sigma^C_J=1-2\sigma^C_1$, we have $\frac{1}{3}>\sigma^C_1$.
Then Rank-order applied to Row implies $\sigma^R_1< \sigma^R_J$, and so $\sigma^R_1<\frac{1}{3}$.
Contradiction.
Suppose $\sigma^R_1<\frac{1}{3}$. Then Rank-order applied to Col implies $\sigma^C_J < \sigma^C_1$, and so $\frac{1}{3}<\sigma^C_1$.
Then Rank-order applied to Row implies $\sigma^R_1> \sigma^R_J$, and so $\sigma^R_1>\frac{1}{3}$.
Contradiction.
Therefore $\sigma^R_1=\frac{1}{3}$, and hence $\sigma^R=\sigma^C=(\frac{1}{3},\frac{1}{3},\frac{1}{3})$ in \emph{any} regular QRE, so the equilibrium is unique.

Consider Game 2. Expected utility of Row in this game from choosing each of her three pure strategies is
\begin{IEEEeqnarray*}{rCl}
  u_{R1}(\mathbf{\sigma}^C)&=& 10+20\sigma^C_2 \\
  u_{R2}(\mathbf{\sigma}^C)&=& 10+20\sigma^C_1 \\
  u_{RJ}(\mathbf{\sigma}^C)&=& 55-45\sigma^C_1-45\sigma^C_2
\end{IEEEeqnarray*}
The previous analysis immediately implies that in \emph{any} regular QRE, $\sigma^R_1=\sigma^R_2$, and $\sigma^C_1=\sigma^C_2$. Suppose $\sigma^R_1\geq \frac{1}{3}$.  Then Rank-order applied to Col implies $\sigma^C_1 \leq \sigma^C_J$, and since  $\sigma^C_J=1-2\sigma^C_1$, we have $\sigma^C_1\leq \frac{1}{3}$. Then Rank-order applied to Row implies $\sigma^R_1<\sigma^R_J$, so $\sigma^R_1<\frac{1}{3}$. Contradiction.
Therefore, $\sigma^R_1<\frac{1}{3}$.  Then Rank-order applied to Col implies $\sigma^C_1 > \sigma^C_J$, and since  $\sigma^C_J=1-2\sigma^C_1$, we have $\sigma^C_1>\frac{1}{3}$. If we also had $\sigma^C_1>\frac{9}{22}$, then Rank-order applied to Row would imply $\sigma^R_1>\sigma^R_J$, hence $\sigma^R_1>\frac{1}{3}$, contradiction.  Thus in any regular QRE, $\frac{1}{3}<\sigma^C_1\leq \frac{9}{22}$ and $\sigma^R_1<\frac{1}{3}$.
It remains to prove that $\sigma^R_1$ and $\sigma^C_1$ are uniquely defined.
Applying Responsiveness to Col implies that $\sigma^C_1$ is strictly increasing in $U_{C1}$, and therefore is strictly decreasing in $\sigma^R_1$. Using the same argument for Row, $\sigma^R_1$ is strictly increasing in $\sigma^C_1$. Therefore any regular QRE in Game 2 is unique.

Consider Game 3. Expected utility of Row in this game from choosing each of her three pure strategies is
\begin{IEEEeqnarray*}{rCl}
  u_{R1}(\mathbf{\sigma}^C)&=& 10+15\sigma^C_1+20\sigma^C_2 \\
  u_{R2}(\mathbf{\sigma}^C)&=& 10+20\sigma^C_1+15\sigma^C_2 \\
  u_{RJ}(\mathbf{\sigma}^C)&=& 30-20\sigma^C_1-20\sigma^C_2
\end{IEEEeqnarray*}
As before, it is easy to show that in \emph{any} regular QRE, $\sigma^R_1=\sigma^R_2$, and therefore $\sigma^C_1=\sigma^C_2$. Applying Responsiveness to Col implies that $\sigma^C_1$ is strictly increasing in $U_{C1}$, and therefore is strictly decreasing in $\sigma^R_1$. Using the same argument for Row, $\sigma^R_1$ is strictly increasing in $\sigma^C_1$. Therefore any regular QRE in Game 3 is unique.
To prove the bounds on QRE probabilities, suppose
$\sigma^R_1\leq \frac{1}{3}$. Then Rank-order applied to Col implies $\sigma^C_1 \geq \sigma^C_J$, and since  $\sigma^C_J=1-2\sigma^C_1$, we have $\sigma^C_1\geq \frac{1}{3}$. Then Rank-order applied to Row implies
$\sigma^R_1>\sigma^R_J$, so $\sigma^R_1>\frac{1}{3}$. Contradiction.
Therefore, $\sigma^R_1>\frac{1}{3}$. Then Rank-order applied to Col implies $\sigma^C_1 < \sigma^C_J$, and since  $\sigma^C_J=1-2\sigma^C_1$, we have $\sigma^C_1<\frac{1}{3}$.  If we also had $\sigma^C_1<\frac{4}{15}$, then Rank-order applied to Row would imply $\sigma^R_1<\sigma^R_J$, hence $\sigma^R_1<\frac{1}{3}$, contradiction.  Thus in any regular QRE, $\frac{4}{15}\leq \sigma^C_1< \frac{1}{3}$ and $\sigma^R_1>\frac{1}{3}$.

Finally, consider Game 4. We will now write $\sigma^C_2=1-\sigma^C_1-\sigma^C_J$, then the expected utility of Row in this game from choosing each of her three pure strategies is
\begin{IEEEeqnarray*}{rCl}
  u_{R1}(\mathbf{\sigma}^C)&=& 30-10\sigma^C_1-20\sigma^C_J \\
  u_{R2}(\mathbf{\sigma}^C)&=& 10+20\sigma^C_1\\
  u_{RJ}(\mathbf{\sigma}^C)&=& 10+20\sigma^C_J
\end{IEEEeqnarray*}

Consider Col's equilibrium strategy. There are two possibilities: 1) $\sigma^C_1> \sigma^C_J$. Then Rank-order applied to Row implies     $\sigma^R_2> \sigma^R_J$, hence $\sigma^R_1+2\sigma^R_2>1$.
    Then Rank-order applied to Col implies $\sigma^C_1< \sigma^C_J$. Contradiction. 2) $\sigma^C_1< \sigma^C_J$. Then Rank-order applied to Row implies
    $\sigma^R_2< \sigma^R_J$, hence  $\sigma^R_1+2\sigma^R_2<1$. Then Rank-order applied to Col implies
    $\sigma^C_1> \sigma^C_J$. Contradiction.
Therefore, in \emph{any} regular QRE in Game 4, $\sigma^C_1=\sigma^C_J$, and consequently, $\sigma^R_2=\sigma^R_J$ (or, equivalently, $\sigma^R_1+2\sigma^R_2=1$).
Applying Responsiveness to Row, $\sigma^R_J$ is strictly increasing in $U_{RJ}$, and therefore is strictly increasing in $\sigma^C_J\equiv \sigma^C_1$. Using the same argument for Col, $\sigma^C_1$ is strictly decreasing in $\sigma^R_2\equiv \sigma^R_J$. Therefore any regular QRE in Game 4 is unique.
To prove the bounds on QRE probabilities, suppose
$\sigma^R_2\geq \sigma^R_1$. Then by Rank-order applied to Col, $\sigma^C_1 \leq \sigma^C_2$ and so $\sigma^C_1\leq \frac{1}{3}$. But then Rank-order applied to Row implies $\sigma^R_2<\sigma^R_1$. Contradiction.
Hence $\sigma^R_2 < \sigma^R_1$, and so $\sigma^R_2<\frac{1}{3}$. Then Rank-order applied to Col implies $\sigma^C_1 > \sigma^C_2$, and so $\sigma^C_1>\frac{1}{3}$. If $\sigma^C_1>\frac{2}{5}$, then by Rank-order $\sigma^R_1< \sigma^R_2$. Contradiction. Therefore $\frac{1}{3}<\sigma^C_1\leq \frac{2}{5}$ and $\sigma^R_2<\frac{1}{3}$. \hfill $\square$
\end{proof}

\section{Additional Details for Computing the Test Statistic}\label{appxC}
As defined in the main text, the $P$-dimensional vector $\bm\nu$
contains the value of the CM inequalities evaluated at the choice
frequencies observed in the experimental data.  Specifically,
the $\ell$-th component of $\bm\nu$, corresponding to a given cycle $G_0,\ldots, G_{\mathcal{L}}$ of games is given by
\begin{IEEEeqnarray*}{rCl}
\nu_\ell&=&\sum_{m=G_0}^{G_\mathcal{L}} \pi_{i1}^m \left[\pi_{k1}^{m+1} u_{i}^{m+1}(s_{i1}, s_{k1})-  \pi_{k1}^{m} u_{i}^{m}(s_{i1},s_{k1}) + \pi_{k2}^{m+1} u_{i}^{m+1}(s_{i1}, s_{k2})- \pi_{k2}^{m} u_{i}^{m}(s_{i1},s_{k2}) \right.\\
&+& \left.(1-\pi_{k1}^{m+1}-\pi_{k2}^{m+1}) u_{i}^{m+1}(s_{i1},s_{kJ})- (1-\pi_{k1}^{m}-\pi_{k2}^{m}) u_{i}^{m}(s_{i1},s_{kJ}) \right] \\
&+& \pi_{i2}^m \left[\pi_{k1}^{m+1} u_{i}^{m+1}(s_{i2},s_{k1})- \pi_{k1}^{m} u_{i}^{m}(s_{i2},s_{k1})
+ \pi_{k2}^{m+1} u_{i}^{m+1}(s_{i2}, s_{k2})- \pi_{k2}^{m} u_{i}^{m}(s_{i2},s_{k2}) \right.\\
&+& \left.(1-\pi_{k1}^{m+1}-\pi_{k2}^{m+1}) u_{i}^{m+1}(s_{i2},s_{kJ})- (1-\pi_{k1}^{m}-\pi_{k2}^{m}) u_{i}^{m}(s_{i2},s_{kJ}) \right] \\
&+&(1-\pi_{i1}^m-\pi_{i2}^m) \left[\pi_{k1}^{m+1} u_{i}^{m+1}(s_{iJ},s_{k1})- \pi_{k1}^{m} u_{i}^{m}(s_{iJ},s_{k1}) \right.\\
&+& \pi_{k2}^{m+1} u_{i}^{m+1}(s_{iJ}, s_{k2})- \pi_{k2}^{m} u_{i}^{m}(s_{iJ},s_{k2}) \\
&+& \left.(1-\pi_{k1}^{m+1}-\pi_{k2}^{m+1}) u_{i}^{m+1}(s_{iJ},s_{kJ})- (1-\pi_{k1}^{m}-\pi_{k2}^{m}) u_{i}^{m}(s_{iJ},s_{kJ}) \right] \\
\end{IEEEeqnarray*}
where we use $i$ to denote the Row player, $k$ to denote the Column player, and $\ell$ changes from $1$ to $20$. For the Column player and $\ell\in[21,40]$ the analogous expression is as follows:
\begin{IEEEeqnarray*}{rCl}
\nu_\ell&=& \sum_{m=G_0}^{G_\mathcal{L}} \pi_{k1}^m \left[\pi_{i1}^{m+1} u_{k}^{m+1}(s_{i1}, s_{k1})-  \pi_{i1}^{m} u_{k}^{m}(s_{i1},s_{k1}) + \pi_{i2}^{m+1} u_{k}^{m+1}( s_{i2}, s_{k1})- \pi_{i2}^{m} u_{k}^{m}(s_{i2}, s_{k1}) \right.\\
&+& \left.(1-\pi_{i1}^{m+1}-\pi_{i2}^{m+1}) u_{k}^{m+1}(s_{iJ}, s_{k1})- (1-\pi_{i1}^{m}-\pi_{i2}^{m}) u_{k}^{m}(s_{iJ}, s_{k1}) \right] \\
&+& \pi_{k2}^m \left[\pi_{i1}^{m+1} u_{k}^{m+1}(s_{i1}, s_{k2})- \pi_{i1}^{m} u_{k}^{m}(s_{i1},s_{k2})
+ \pi_{i2}^{m+1} u_{k}^{m+1}(s_{i2}, s_{k2})- \pi_{i2}^{m} u_{k}^{m}(s_{i2},s_{k2}) \right.\\
&+& \left.(1-\pi_{i1}^{m+1}-\pi_{i2}^{m+1}) u_{k}^{m+1}(s_{iJ}, s_{k2})- (1-\pi_{i1}^{m}-\pi_{i2}^{m}) u_{k}^{m}(s_{iJ}, s_{k2}) \right] \\
&+&(1-\pi_{k1}^m-\pi_{k2}^m) \left[\pi_{i1}^{m+1} u_{k}^{m+1}(s_{i1}, s_{kJ})- \pi_{i1}^{m} u_{k}^{m}(s_{i1},s_{kJ}) \right.\\
&+& \pi_{i2}^{m+1} u_{k}^{m+1}(s_{i2}, s_{kJ})- \pi_{i2}^{m} u_{k}^{m}(s_{i2},s_{kJ}) \\
&+& \left.(1-\pi_{i1}^{m+1}-\pi_{i2}^{m+1}) u_{k}^{m+1}(s_{iJ},s_{kJ})- (1-\pi_{i1}^{m}-\pi_{i2}^{m}) u_{k}^{m}(s_{iJ},s_{kJ}) \right] \\
\end{IEEEeqnarray*}
We differentiate the above expressions with respect to $\pi^{m}$ to obtain a $P\times 16$ estimate of the Jacobian $\hat{J}= \frac{\partial}{\partial \bm{\pi}}\bm{\mu}(\bm{\hat{\pi}})$ in order to compute an estimate of the variance-covariance matrix $\hat{\Sigma}_{[P\times P]}=\hat{J}\hat{V}\hat{J}'$ by the Delta method.
For the case of four games, the partial derivatives form the $40\times 16$ matrix $\hat{J}$. The first $20$ rows correspond to the differentiated LHS of the cycles for the Row player, and the last $20$ rows correspond to the differentiated LHS of the cycles for the Column player. The first $8$ columns correspond to the derivatives with respect to $\pi_{i1}^m$, $\pi_{i2}^m$, and the last $8$ columns correspond to the derivatives with respect to $\pi_{k1}^m$, $\pi_{k2}^m$, $m\in\{1,\ldots,4\}$.\footnote{Clearly, the probability to choose Joker can be expressed via the probabilities to choose 1 and 2, using the total probability constraint.}

Let $S_0^m\equiv \{\ell\in\{1,\ldots,40\}|m\not\in C_\ell \}$ be the set of cycle indices such that corresponding cycles (in the order given in \eqref{ordered_cycles}) do not include game $m$. E.g., for $m=1$, $S_0^m=\{4,5,6, 13,14, 24,25,26,33,34\}$.
Let $S_i^m\equiv \{\ell\in\{1,\ldots,20\}|\ell \not\in S_0^m \}$ be a subset of cycle indices that include game $m$ and pertain to the Row player, and let $S_k^m\equiv \{\ell\in\{21,\ldots,40\}|\ell \not\in S_0^m \}$ be a subset of cycle indices that include game $m$ and pertain to the Column player.
Finally, for a cycle of length $\mathcal{L}$, denote $\ominus\equiv - \mod \mathcal{L}$ subtraction modulus $\mathcal{L}$.

We can now express the derivatives with respect to $\pi_{i1}^m$ and $\pi_{i2}^m$, $m\in\{1,\ldots,4\}$, in the following general form. The partial derivatives wrt $\pi_{i1}^m$ are
\begin{IEEEeqnarray*}{+rCl+x*}
\frac{\partial\nu_\ell}{\partial\pi_{i1}^m}&=& 0 \hfill & for $\ell\in S_0^{m}$\\
\frac{\partial\nu_\ell}{\partial\pi_{i1}^m}&=&
\pi_{k1}^{m+1} u_{i}^{m+1}(s_{i1}, s_{k1})-  \pi_{k1}^{m} u_{i}^{m}(s_{i1},s_{k1}) + \pi_{k2}^{m+1} u_{i}^{m+1}(s_{i1}, s_{k2}) & \\
&-& \pi_{k2}^{m} u_{i}^{m}(s_{i1},s_{k2}) + (1-\pi_{k1}^{m+1}-\pi_{k2}^{m+1}) u_{i}^{m+1}(s_{i1},s_{kJ})- (1-\pi_{k1}^{m}-\pi_{k2}^{m}) u_{i}^{m}(s_{i1},s_{kJ}) & \\
&-& \left[\pi_{k1}^{m+1} u_{i}^{m+1}(s_{iJ},s_{k1})- \pi_{k1}^{m} u_{i}^{m}(s_{iJ},s_{k1}) + \pi_{k2}^{m+1} u_{i}^{m+1}(s_{iJ}, s_{k2})- \pi_{k2}^{m} u_{i}^{m}(s_{iJ},s_{k2}) \right.&\\
&+& \left.(1-\pi_{k1}^{m+1}-\pi_{k2}^{m+1}) u_{i}^{m+1}(s_{iJ},s_{kJ})- (1-\pi_{k1}^{m}-\pi_{k2}^{m}) u_{i}^{m}(s_{iJ},s_{kJ}) \right] \hfill &  for $\ell\in S_{i}^m$ \\
\frac{\partial\nu_\ell}{\partial\pi_{i1}^m}&=& \pi_{k1}^m \left[- u_{k}^{m}(s_{i1},s_{k1}) + u_{k}^{m}(s_{iJ}, s_{k1}) \right] + \pi_{k2}^m \left[- u_{k}^{m}(s_{i1},s_{k2}) +  u_{k}^{m}(s_{iJ}, s_{k2}) \right] & \\
&+&(1-\pi_{k1}^m-\pi_{k2}^m) \left[- u_{k}^{m}(s_{i1},s_{kJ})+ u_{k}^{m}(s_{iJ},s_{kJ}) \right] & \\
&+&\pi_{k1}^{m\ominus1} \left[u_{k}^{m}(s_{i1}, s_{k1}) - u_{k}^{m}(s_{iJ}, s_{k1}) \right] + \pi_{k2}^{m\ominus1} \left[u_{k}^{m}(s_{i1}, s_{k2})-u_{k}^{m}(s_{iJ}, s_{k2}) \right] & \\
&+&(1-\pi_{k1}^{m\ominus1}-\pi_{k2}^{m\ominus1}) \left[u_{k}^{m}(s_{i1}, s_{kJ})- u_{k}^{m}(s_{iJ},s_{kJ}) \right] \hfill &  for $\ell\in S_{k}^m$\\
\end{IEEEeqnarray*}
The partial derivatives wrt $\pi_{i2}^m$ are
\begin{IEEEeqnarray*}{+rCl+x*}
\frac{\partial\nu_\ell}{\partial\pi_{i2}^m}&=& 0 \hfill & for $\ell\in S_0^m$\\
\frac{\partial\nu_\ell}{\partial\pi_{i2}^m}&=&
 \left[\pi_{k1}^{m+1} u_{i}^{m+1}(s_{i2},s_{k1})- \pi_{k1}^{m} u_{i}^{m}(s_{i2},s_{k1})
+ \pi_{k2}^{m+1} u_{i}^{m+1}(s_{i2}, s_{k2})- \pi_{k2}^{m} u_{i}^{m}(s_{i2},s_{k2}) \right. & \\
&+& \left.(1-\pi_{k1}^{m+1}-\pi_{k2}^{m+1}) u_{i}^{m+1}(s_{i2},s_{kJ})- (1-\pi_{k1}^{m}-\pi_{k2}^{m}) u_{i}^{m}(s_{i2},s_{kJ}) \right] & \\
&-&\left[\pi_{k1}^{m+1} u_{i}^{m+1}(s_{iJ},s_{k1})- \pi_{k1}^{m} u_{i}^{m}(s_{iJ},s_{k1}) + \pi_{k2}^{m+1} u_{i}^{m+1}(s_{iJ}, s_{k2})- \pi_{k2}^{m} u_{i}^{m}(s_{iJ},s_{k2}) \right. &\\
&+& \left.(1-\pi_{k1}^{m+1}-\pi_{k2}^{m+1}) u_{i}^{m+1}(s_{iJ},s_{kJ})- (1-\pi_{k1}^{m}-\pi_{k2}^{m}) u_{i}^{m}(s_{iJ},s_{kJ}) \right] \hfill &  for $\ell\in S_{i}^m$ \\
\frac{\partial\nu_\ell}{\partial\pi_{i2}^m}&=&
\pi_{k1}^m \left[- u_{k}^{m}(s_{i2}, s_{k1}) +  u_{k}^{m}(s_{iJ}, s_{k1}) \right] + \pi_{k2}^m \left[-u_{k}^{m}(s_{i2},s_{k2})+ u_{k}^{m}(s_{iJ}, s_{k2}) \right] & \\
&+&(1-\pi_{k1}^m-\pi_{k2}^m) \left[- u_{k}^{m}(s_{i2},s_{kJ}) + u_{k}^{m}(s_{iJ},s_{kJ}) \right] &\\
&+& \pi_{k1}^{m\ominus1} \left[u_{k}^{m}( s_{i2}, s_{k1})- u_{k}^{m}(s_{iJ}, s_{k1})\right] + \pi_{k2}^{m\ominus1} \left[ u_{k}^{m}(s_{i2}, s_{k2})-u_{k}^{m}(s_{iJ}, s_{k2})\right] & \\
&+&(1-\pi_{k1}^{m\ominus1}-\pi_{k2}^{m\ominus1}) \left[ u_{k}^{m}(s_{i2}, s_{kJ})- u_{k}^{m}(s_{iJ},s_{kJ}) \right] \hfill &  for $\ell\in S_{k}^m$
\end{IEEEeqnarray*}

To obtain the derivatives with respect to $\pi_{k1}^m$ and $\pi_{k2}^m$, one just needs to use the corresponding partial derivatives wrt $\pi_{i1}^m$ and $\pi_{i2}^m$, and exchange everywhere the subscripts $i$ and $k$, so we omit the derivation. For the sake of completenes, though, we list the indices subsets for each game $m\in\{1,..,4\}$ in Table \ref{Table:cyc_indices}.
\begin{table}[htbp]\centering\caption{Sets of cycle indices for each game.}\label{Table:cyc_indices}
\begin{center} \scalebox{0.9}{
\begin{tabularx}{\textwidth}{lXXX}\hline\hline
$m$ & $S_0^m$ & $S_i^m$  & $S_k^m$ \\ \hline
1 & 4, 5, 6, 13, 14,\newline 24, 25, 26, 33, 34 & 1, 2, 3, 7, 8,\newline 9, 10, 11, 12, 15,\newline 16, 17, 18, 19, 20 & 21, 22, 23, 27, 28,\newline 29, 30, 31, 32, 35,\newline 36, 37, 38, 39, 40 \\ \hline
2 & 2, 3, 6, 10, 12, \newline 22, 23, 26, 30, 32 & 1, 4, 5, 7, 8, \newline 9, 11, 13, 14, 15, \newline 16, 17, 18, 19, 20 & 21, 24, 25, 27, 28,\newline 29, 31, 33, 34, 35,\newline 36, 37, 38, 39, 40 \\ \hline
3 & 1, 3, 5, 8, 11, \newline 21, 23, 25, 28, 31 & 2, 4, 6, 7, 9, \newline 10, 12, 13, 14, 15, \newline 16, 17, 18, 19, 20 & 22, 24, 26, 27, 29,\newline 30, 32, 33, 34, 35,\newline 36, 37, 38, 39, 40 \\ \hline
4 & 1, 2, 4, 7, 9 \newline 21, 22, 24, 27, 29 & 3, 5, 6, 8, 10, \newline 11, 12, 13, 14, 15, \newline 16, 17, 18, 19, 20 & 23, 25, 26, 28, 30,\newline 31, 32, 33, 34, 35,\newline 36, 37, 38, 39, 40 \\ \hline \hline
\multicolumn{4}{p{0.8\textwidth}}{\footnotesize \emph{Notes.} The cycle indices are for the Row player. To obtain the corresponding Column player cycle indices, swap the last two columns. }
\end{tabularx}}
\end{center}
\end{table}

\section{Experiment Instructions }\label{appx}
The instructions in the experiment, given below, largely follow \citet{McKelveyPalfreyWeber00}.

This is an experiment in decision making, and you will be paid for your participation in cash. Different subjects may earn different amounts. What you earn depends partly on your decisions and partly on the decisions of others.

The entire experiment will take place through computer terminals, and all interaction between subjects will take place through the computers. It is important that you do not talk or in any way try to communicate with other subjects during the experiment. If you violate the rules, we may ask you to leave the experiment.

We will start with a brief instruction period. If you have any questions during the instruction period, raise your hand and your question will be answered so everyone can hear. If any difficulties arise after the experiment has begun, raise your hand, and an experimenter will come and assist you.

This experiment consists of several periods or matches and will take between $30$ to $60$ minutes. I will now describe what occurs in each match.

[Turn on the projector]

First, you will be randomly paired with another subject, and each of you will simultaneously be asked to make a choice.

Each subject in each pair will be asked to choose one of the three rows in the table which will appear on the computer screen, and which is also shown now on the screen at the front of the room. Your choices will be always displayed as rows of this table, while your partner's choices will be displayed as columns. It will be the other way round for your partner: for them, your choices will be displayed as columns, and their choices as rows.

You can choose the first, the second, or the third row. Neither you nor your partner will be informed of what choice the other has made until after all choices have been made.

After each subject has made his or her choice, payoffs for the match are determined based on the choices made. Payoffs to you are indicated by the red numbers in the table, while payoffs to your partner are indicated by the blue numbers. Each cell represents a pair of payoffs from your choice and the choice of your partner. The units are in francs, which will be exchanged to US dollars at the end of the experiment.

For example, if you choose 'A' and your partner chooses 'D', you receive a payoff of 10 francs, while your partner receives a payoff of $20$ francs. If you choose 'A' and your partner chooses 'F', you receive a payoff of $30$ francs, while your partner receives a payoff of $30$ francs. If you choose 'C' and your partner chooses 'E', you receive a payoff of $10$ francs, while your partner receives a payoff of $20$ francs. And so on.

Once all choices have been made the resulting payoffs and choices are displayed, the history panel is updated and the match is completed.

[show the slide with a completed match]

This process will be repeated for several matches. The end of the experiment will be announced without warning. In every match, you will be randomly paired with a new subject. The identity of the person you are paired with will never be revealed to you. The payoffs and the labels may change every match.

After some matches, we will ask you to indicate what you think is the likelihood that your current partner has made a particular choice. This is what it looks like.

[show slide with belief elicitation]

Suppose you think that your partner has a 15\% chance of choosing 'D' and a 60\% chance of choosing 'E'. Indicate your opinion using the slider, and then press 'Confirm'. Once all subjects have indicated their opinions and confirmed them, the resulting payoffs and choices are displayed, the history panel is updated and the match is completed as usual.

Your final earnings for the experiment will be the sum of your payoffs from all matches. This amount in francs will be exchanged into U.S. dollars using the exchange rate of $90$ cents for $100$ francs. You will see your total payoff in dollars at the end of the experiment. You will also receive a show-up fee of \$7.
Are there any questions about the procedure?

[wait for response]

We will now start with four practice matches. Your payoffs from the practice matches are not counted in your total. In the first three matches you will be asked to choose one of the three rows of a table. In the fourth match you will be also asked to indicate your opinion about the likelihood of your partner's choices for each of three actions. Is everyone ready?

[wait for response]

Now please double click on the 'Client Multistage' icon on your desktop. The program will ask you to type in your name. Please type in the number of your computer station instead.

[wait for subjects to connect to server]

We will now start the practice matches. Do not hit any keys or click the mouse button until you are told to do so.

[start first practice match]

You see the experiment screen. In the middle of the screen is the table which you have previously seen up on the screen at the front of the room. At the top of the screen, you see your subject ID number, and your computer name. You also see the history panel which is currently empty.

We will now start the first practice match. Remember, do not hit any keys or click the mouse buttons until you are told to do so. You are all now paired with someone from this class and asked to choose one of the three rows. Exactly half of your see label 'A' at the left hand side of the top row, while the remaining half now see label 'D' at the same row.

Now, all of you please move the mouse so that it is pointing to the top row. You will see that the row is highlighted in red. Move the mouse to the bottom row – and the highlighting goes along with the mouse. To choose a row you just click on it. Now please click once anywhere on the bottom row.

[Wait for subjects to move mouse to appropriate row]

After all subjects have confirmed their choices, the match is over. The outcome of this match, 'C'–-'F', is now highlighted on everybody's screen. Also, note that the moves and payoffs of the match are recorded in the history panel. The outcomes of all of your previous matches will be recorded there throughout the experiment so that you can refer back to previous outcomes whenever you like. The payoff to the subject who chose 'C' for this match is $20$, and the payoff to the subject who chose $F$ is '10'.

You are not being paid for the practice session, but if this were the actual experiment, then the payoff you see on the screen would be money (in francs) you have earned from the first match. The total you earn over all real matches, in addition to the show-up fee, is what you will be paid for your participation in the experiment.

We will now proceed to the second practice match.

[Start second match]

For the second match, you have been randomly paired with a different subject. You are not paired with the same person you were paired with in the first match. The rules for the second match are exactly like for the first.  Please make your choices.

[Wait for subjects]

We will now proceed to the third practice match. The rules for the third match are exactly like the first.  Please make your choices.

[Start third match]

We will now proceed to the fourth practice match. The rules for the fourth match are exactly like the first. Please make your choices.

[Wait for subjects]

Now that you have made your choice, you see that a slider appears asking you to indicate the relative likelihood of your partner choosing each of the available actions. There is also a confirmation button. Please indicate your opinion by adjusting the thumbs and then press 'Confirm'.

[wait for subjects]

This is the end of the practice match. Are there any questions?

[wait for response]

Now let's start the actual experiment. If there are any problems from this point on, raise your hand and an experimenter will come and assist you. Please pull up the dividers between your cubicles.

[start the actual session]

The experiment is now completed. Thank you all very much for participating in this experiment.
Please record your total payoff from the matches in U.S. dollars at the experiment record sheet. Please add your show-up fee and write down the total, rounded up to the nearest dollar. After you are done with this, please remain seated. You will be called by your computer name and paid in the office at the back of the room one at a time. Please bring all your things with you when you go to the back office. You can leave the experiment through the back door of the office.

Please refrain from discussing this experiment while you are waiting to receive payment so that privacy regarding individual choices and payoffs may be maintained.

\end{document}